\documentclass[11pt]{elsarticle}

\usepackage{lineno,hyperref}

\usepackage{float}
\usepackage{graphicx}
\usepackage{subfig}
\usepackage{mathrsfs}
\usepackage{graphics}
\usepackage{float}
\usepackage{amsmath}
\usepackage{amsfonts}
\usepackage{mathrsfs}
\usepackage{amsmath}
\usepackage{longtable}
\usepackage{optidef}
\usepackage[subnum]{cases}

\usepackage{optidef}
\usepackage{graphicx}
\usepackage{float}
\usepackage{graphicx}
\usepackage{subfig}
\usepackage{mathrsfs}
\usepackage{float}
\usepackage{amsmath}
\usepackage{amsfonts}
\usepackage{booktabs}
\usepackage{natbib}
\usepackage{bbm}
\usepackage{algorithm}
\usepackage[noend]{algpseudocode}
\usepackage{multirow}
\algdef{SE}[DOWHILE]{Do}{doWhile}{\algorithmicdo}[1]{\algorithmicwhile\ #1}

\usepackage[letterpaper, portrait, margin=1in]{geometry}

\usepackage{booktabs}
\usepackage{tabularx}

\usepackage{xcolor}
\hypersetup{
    colorlinks,
    linkcolor={red!50!red},
    citecolor={blue!50!blue},
    urlcolor={blue!80!blue}
}

\usepackage{optidef}

\usepackage{amsthm}

\theoremstyle{definition}

\usepackage{newtxtext} 

\newcommand{\One}{\mathbbm{1}}

\makeatletter
\newcommand*{\centerfloat}{%
  \parindent \z@
  \leftskip \z@ \@plus 1fil \@minus \marginparwidth
  \rightskip \leftskip
  \parfillskip \z@skip}
\makeatother

\usepackage{amssymb,amsmath,caption}
\usepackage[hang,flushmargin]{footmisc}
\newcommand{\algorithmfootnote}[2][\footnotesize]{
  \let\old@algocf@finish\@algocf@finish
  \def\@algocf@finish{\old@algocf@finish
    \leavevmode\rlap{\begin{minipage}{\linewidth}
    #1#2
    \end{minipage}}%
  }%
}

\usepackage{setspace}
\usepackage{natbib}
\usepackage{algorithm}
\usepackage[noend]{algpseudocode}
\algdef{SE}[DOWHILE]{Do}{doWhile}{\algorithmicdo}[1]{\algorithmicwhile\ #1}

\usepackage{appendix}

\usepackage{array}
\usepackage{threeparttable}
\usepackage{multirow}
\usepackage{rotating}
\usepackage{makecell}

\journal{}

\biboptions{authoryear}

\begin{document}
\begin{frontmatter}

\title{Modeling virus transmission risks in commuting with emerging mobility services: A case study of COVID-19}


\author[label1]{Baichuan Mo\corref{mycorrespondingauthor}}
\author[label4]{Peyman Noursalehi}
\author[label3]{Haris N. Koutsopoulos}
\author[label4]{Jinhua Zhao}
\address[label1]{Department of Civil and Environmental Engineering, Massachusetts Institute of Technology, Cambridge, MA 02139}
\address[label4]{Department of Urban Studies and Planning, Massachusetts Institute of Technology, Cambridge, MA 20139}
\address[label3]{Department of Civil and Environmental Engineering, Northeastern University, Boston, MA 02115}

\cortext[mycorrespondingauthor]{Corresponding author}

\begin{abstract}
Commuting is an important part of daily life. With the gradual recovery from COVID-19 and more people returning to work from the office, the transmission of COVID-19 during commuting becomes a concern. Recent emerging mobility services (such as ride-hailing and bike-sharing) further deteriorate the infection risks due to shared vehicles or spaces during travel. Hence, it is
important to quantify the infection risks in commuting. This paper proposes a probabilistic framework to estimate the risk of infection during an individual's commute considering different travel modes, including public transit, ride-share, bike, and walking. The objective is to evaluate the probability of infection as well as the estimation errors (i.e., uncertainty quantification) given the origin-destination (OD), departure time, and travel mode. We first define a general trip planning function to generate trip trajectories and probabilities of choosing different paths according to the OD, departure time, and travel mode. Then, we consider two channels of infections: 1) infection by close contact and 2) infection by touching surfaces. The infection risks are calculated on a trip segment basis. Different sources of data (such as smart card data, travel surveys, and population data) are used to estimate the potential interactions between the individual and the infectious environment. A first-order approximation is used to simplify the computational complexity. We also derive the closed-form formulation for the estimation errors, enabling us to quantify the uncertainty of the estimation results. The model is implemented in the MIT community as a case study. We evaluate the commute infection risks for employees and students. Results show that most of the individuals have an infection probability close to zero. The maximum infection probability is around 0.8\%, implying that the probability of getting infected during the commuting process is low. Individuals with larger travel distances, traveling in transit, and traveling during peak hours are more likely to get infected. Practical implementations of the model are also discussed.  
\end{abstract}


\end{frontmatter}


\section{Introduction}\label{intro}
COVID-19 has greatly affected people's lives all over the world. Recently, with the vaccination and people's prevention consciousness, we are stepping into a new era of living with the virus. With the gradual recovery from the pandemic, more and more people return to work from the office. 

Commuting is an important part of the daily lives of people working in an office. In light of the infectiousness of COVID-19, the infection risk during the commuting process is a concern, especially for people using public transportation, as indicated by many previous studies \citep{mo2021modeling, zhou2021virus}. On the other hand, recent emerging mobility services (such as ride-hailing and bike-sharing) further deteriorate the infection risks due to shared vehicles or spaces during travel. Therefore, it is important to quantify the infection risks in commuting more broadly, where the results are helpful for people to evaluate their health risks and better inform their commuting route/travel mode choices, and for policymakers to reach informed decisions. 

Many previous studies have modeled the COVID-19 infection risks in public transit systems, an important travel mode of commuting. These studies can be categorized from the macro-level at the city scale \citep{mo2021modeling} or the micro-scale at the vehicle scale \citep{shinohara2021survey, zhou2021virus}. Researchers have also considered the impact of commuting on the broader spatial transmission of COVID-19 and the related control strategies \citep{mitze2022propagation,ando2021effect, fajgelbaum2021optimal, kondo2021simulating}. There are also studies evaluating the impact of COVID-19 on the commuting process with empirical data (such as surveys), such as the impact of COVID-19 on ridership changes, travel mode choices \citep{tan2021choice, medlock2021covid}, and departure time change \citep{ecke2022covid}. 

However, there are still two research gaps. First, none of the previous studies have considered the infection modeling for the commuting process as a whole with multiple travel modes and multi-modal trip itineraries. Second, most of the previous studies regarding infection risk modeling only output the probability of infection (or the $R_0$ value indicating the spreading intensity). The estimation uncertainties (i.e., how accurate are the estimates) are not provided. 

In this study, we propose a probabilistic framework to estimate the risk of infection during an individual's commute considering different travel modes, including public transit, ride-share, bike, and walking. The model enables evaluating both the probability of infection and the estimation errors (i.e., uncertainty quantification). We first define a general trip planning function to generate trip trajectories and probabilities of choosing different paths according to the origin, destination, departure time, and travel mode. Two channels of infection are considered: 1) infection by close contact and 2) infection by touching surfaces. The infection risks are calculated on a trip segment basis. Different sources of data (such as smart card data, travel surveys, and population data) are used to estimate the potential interactions between the individual and the infectious environment. A first-order approximation technique is used to simplify the computational complexity. We derive the closed-form formulations for the estimation errors, enabling us to quantify the uncertainties of the estimation results. The model is implemented in the MIT community as a case study. We evaluate the commute infection risks for employees and students. Results show that most of the individuals have an infection probability close to zero. The maximum infection probability is around 0.8\%, implying that the probability of getting infected during the commuting process is low. Individuals with larger travel distances, traveling with transit, and traveling during peak hours are more likely to get infected.  

The main contribution of this paper is twofold:
\begin{itemize}
    \item This is the first study dedicated to virus transmission modeling during commuting with the consideration of various travel modes and multi-modal trip itineraries. Infections due to close contact and touching surfaces are both captured.
    \item In addition to estimating infection probabilities, this paper also calculates the estimation errors (i.e., the standard deviation of the estimated probabilities) for uncertainty quantification, which has not been done in the literature.  
\end{itemize}

The remainder of the paper is organized as follows. The literature review is shown in Section \ref{sec_liter}. In Section \ref{method}, we describe the problem and discuss the solution methods. We apply the proposed framework to the MIT community as a case study in Section \ref{sec_case_study}. Finally, we conclude our study and summarize the main findings in Section \ref{sec_conclusion}.

\section{Literature review}\label{sec_liter}
\subsection{Infection modeling in public transit}
Public transit is an important travel mode for commuting. Previous studies have explored epidemic spreading and infection risk modeling in transit networks. \citet{mo2021modeling} propose a time-varying weighted encounter network to model the spreading of infectious diseases through public transit systems. The model is implemented at the metropolitan level for population infection calculation. \citet{zhou2021virus} propose a modified Wells-Riley model for infection probability calculation in public transportation systems at the vehicle level. The model captures the spatial and temporal passenger flow characteristics in terms of the number of boarding and alighting passengers and the number of infectors. Similarly, \citet{ku2021safe} analyzed the degree of infection exposure in public transport by simulating how passengers encounter and infect each other during their journeys. \citet{shinohara2021survey} adopted a two-zone-based exponential model to calculate the infection risks in commuter trains by collecting air exchange rate data under various conditions.
\citet{zhao2022method} developed a Wells-Riley model-based method to quantitatively evaluate the infection risk of riding public transit. They compared the effectiveness of different countermeasures in managing the spread. 

\subsection{Impact of COVID-19 on commuting}
COVID-19 may affect the commuting process in many aspects, such as ridership and service frequency decrease, changes in passengers' travel mode choices, route choices, and departure times choices. Previous studies have evaluated these impacts using different sources of empirical data. For example, many studies have used the smart card data to analyze the impact of COVID-19 on transit ridership changes \citep{ahangari2020public, chang2021does, wilbur2020impact, jenelius2020impacts}. There are also studies on ridership changes in ride-hailing systems \citep{meredith2021relationship} and bike-sharing systems \citep{wang2021bikeshare}. \citet{tan2021choice} conducted a survey to understand commuters' mode choice changes during the COVID-19 pandemic. They used a logistic regression model with personal attributes, travel attributes, and perception of COVID-19 based on a  sample of 559 responses to a survey. \citet{ecke2022covid} examine how people's commuting behavior changed before and after COVID-19. The results show that people did not significantly change their commuting behavior in terms of commuting time and commuting mode.

\subsection{Impact of commuting on COVID-19 spreading}
Commuting may contribute to the spreading of COVID-19 by transporting infectious passengers across different regions. For example, \citet{mitze2022propagation} proposed a spatial econometric model of the epidemic spread to identify the role played by commuting-to-work patterns for spatial disease transmission and explored if the imposed containment policies affected the strength of this transmission channel. \citet{ando2021effect} investigated the relationship among commuting, the risk of COVID-19, and COVID-19-induced anxiety using internet-based survey data from 27,036 respondents. \citet{fajgelbaum2021optimal} designed an optimal dynamic lockdown strategy against COVID-19 within a commuting network. \citet{kondo2021simulating} developed a spatial susceptible–exposed–infectious–recovered model to analyze the effects of restricting inter-regional commuting mobility on the spatial spread of the COVID-19 infection in Japan.

\subsection{Research gaps}
To the best of the authors' knowledge, no existing papers have considered dedicated infection
modeling for the commuting process as a whole with multiple travel modes and multi-modal trip itineraries. Most of them focus on infection modeling for a single travel mode (e.g., public transit), or consider the general modeling of epidemic spreading at a city level, where the commuting process is just a part of the big framework without using travel mode or itinerary-specific modeling methodologies. On the other hand, most of the previous studies regarding infection risk modeling only output the probability of infection or the R0 value indicating the spreading intensity. The estimation uncertainties (i.e., how accurate are the results) are not calculated.  

\section{Methodology}\label{method}

\subsection{Problem definition}
Consider a set of individuals $\mathcal{I}$. For each individual $i\in\mathcal{I}$, suppose that we know their origin $o_i$, destination $d_i$, departure time $t_i$, and travel mode $m_i$ for their daily commuting. The objective of this study is to estimate the probability of individual $i$ getting affected: ${\mathbb{P}}(i \text{ infected}\mid o_i,d_i,t_i,m_i)$. Besides, we also aim to quantify the estimation uncertainty. If we treat ${\mathbb{P}}(i \text{ infected}\mid o_i,d_i,t_i,m_i)$ as a random variable, another goal of the study is to obtain the standard error of the estimation: 
$\sqrt{\text{Var}[{\mathbb{P}}(i \text{ infected}\mid o_i,d_i,t_i,m_i)]}$.

In the following sections, we first illustrate how ${\mathbb{P}}(i \text{ infected}\mid o_i,d_i,t_i,m_i)$ is estimated. The estimation of standard errors is illustrated in Section \ref{sec_uncertainty_quan}.



\subsection{Trip itinerary generation}\label{sec_path}
Given an individual $i$’s trip information $(o_i,d_i,t_i,m_i)$, there exists a trip planner function $TP(\cdot)$ that takes $(o_i,d_i,t_i,m_i)$ as input, and outputs a set of feasible paths for the individual $\mathcal{R}_i$ and the associated path choice probability ${\mathbb{P}}_{\text{Path}}(r)$ for all paths $r \in \mathcal{R}_i$, that is:
\begin{align}
   \mathcal{R}_i,\; {\mathbb{P}}_{\text{Path}}(r)_{r\in\mathcal{R}_i}  = TP(o_i,d_i,t_i,m_i)
\end{align}

For example, we can define $TP(\cdot)$ as a composed function of the Google Map API and a C-logit model:
\begin{align}
TP(o_i,d_i,t_i,m_i) = \texttt{C-Logit} \circ \texttt{Google Map API}(o_i,d_i,t_i,m_i)
\end{align}
Specifically, the Google Map API returns the path set $\mathcal{R}_i$ and path attributes $\boldsymbol{X}_r$ for all $r \in \mathcal{R}_i$ (e.g., travel time, travel cost, etc.), that is:
\begin{align}
    \mathcal{R}_i,\; (\boldsymbol{X}_r)_{r\in\mathcal{R}_i} = \texttt{Google Map API}(o_i,d_i,t_i, m_i) 
\end{align}
The C-Logit model outputs the path choice probability and the standard errors for each path. The c-Logit model is an extension of the multinomial logit (MNL) model to correct for the correlation among paths due to overlapping \citep{cascetta1996modified}. The key idea is to define a term called the ``commonality factor'' of path $r$ (i.e., $CF_r$), which measures the degree
of similarity of path $r$ with the other paths of the same OD. Based on the C-logit model, the probability of choosing path r can be calculated as
\begin{align}
    {\mathbb{P}}_{\text{Path}}(r) = \texttt{C-Logit}(\boldsymbol{X}_r) = \frac{\exp[{\boldsymbol{\beta}}^T\cdot(\boldsymbol{X}_r, CF_r)]}{\sum_{r'\in\mathcal{R}_i}\exp[{\boldsymbol{\beta}}^T\cdot(\boldsymbol{X}_{r'}, CF_{r'})]}
\end{align}
where ${\boldsymbol{\beta}}$ is the parameter vector to estimate. The formulation of $CF_r$ can be found in \citet{cascetta1996modified}.

Note the $TP(\cdot)$ can be defined more broadly than google map API plus the C-logit model. Other examples include the k-shortest path in a multiple-modal network compounded with any behavioral model for path choices \citep{mo2022impact}.

\subsection{Infection modeling for a trip segment}\label{sec_inf_seg}
\subsubsection{Definition of a trip segment}
A path $r \in \mathcal{R}_i$ usually contains multiple trip segments, such as walking from home to a bus station, taking a bus, and walking from a bus station to the office. The infection may happen at every trip segment. In this study, we define a segment $s$ of a path as continuous travel with the same travel mode along the path. Let the set of all segments for path $r$ be $\mathcal{S}_r$. It is worth noting that, if a transit trip has one or more transfers, we separate the transit trip into multiple segments based on transfers because the passenger needs to first alight and then board a new vehicle, which is equivalent to changing to a “new travel mode” in infection modeling. We also ignore short walking segments (less than 3 minutes or less than 1km, e.g., transfer walking) for modeling convenience. 

\begin{figure}[H]
\centering
\includegraphics[width = 0.6\linewidth]{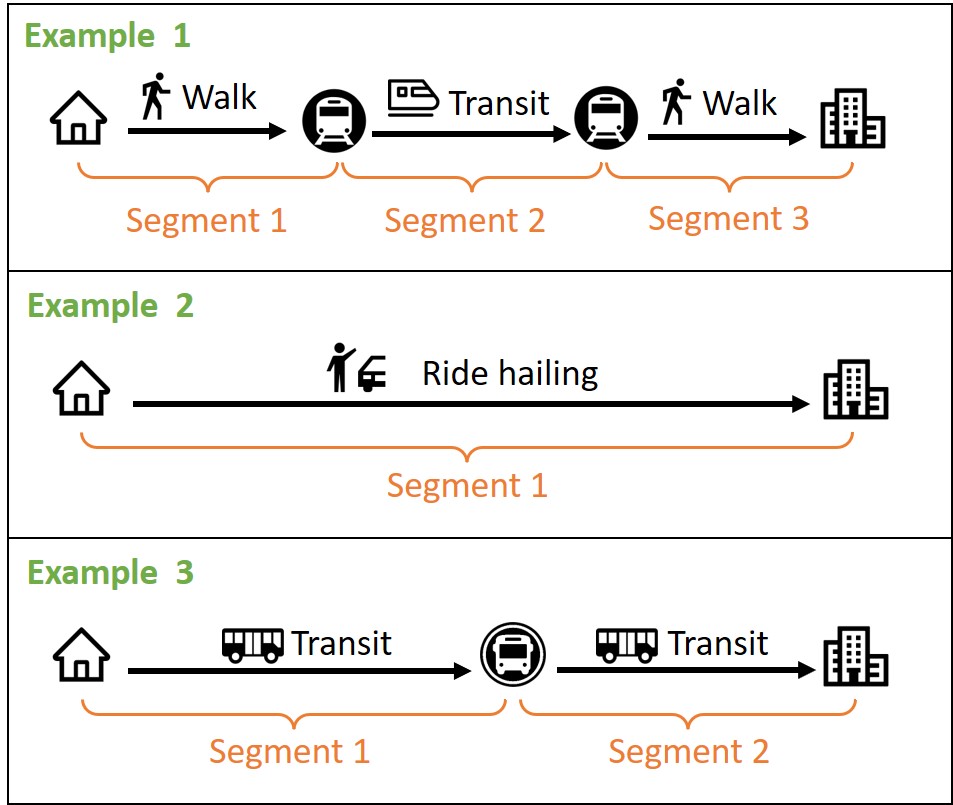}
\caption{Illustration of the segment definition}
\label{fig_seg_example}
\end{figure}

We provide three examples to illustrate the definition of segments (Figure \ref{fig_seg_example}). The first example shows three segments: walking from home to a subway station, taking the subway, and walking from a subway station to the office. The second example is a ride-hailing trip with only a single segment.  The third example shows a transit trip with a transfer, which is separated into two segments by definition. 

\subsubsection{Infection risk for each segment}\label{sec_inf_risk_seg}
We consider two different channels for infection: 1) infected by close contact with infectious persons and 2) infected by touching infectious surfaces. 

\textbf{Infection by close contact}: Consider a trip segment $s \in \mathcal{S}_r$.  We define $\mathcal{P}_s$ as the set of persons that have been in the six feet infectious range of individual $i$.  For a person $p \in \mathcal{P}_s$, let the duration of the interaction between $i$ and $p$ be $d_{i,p}$. If $p$ is infectious, $i$ would have a probability of getting infected. Depending on whether the interaction happens indoors (e.g., in a bus) or outdoors (e.g., walking by), there are two different physical models. 

In an indoor environment, the probability of $i$ getting infected by $p$ can be calculated by the well-known Wells-Riley model \citep{riley1978airborne}:
\begin{align}
    \mathbb{P}_{\text{Indoor}}(p \rightarrow i \mid p \text{ infectious}) = 1 - \exp\left(\frac{-b\cdot q \cdot d_{i,p}}{Q^{\text{Indoor}}}\right)
\end{align}
where $b$ is the  breathing rate per person ($m^3$ /hour); $q$ is the quanta generation rate (/hour),  $Q$ is the room ventilation rate of clean air ($m^3$ /hour).

In an outdoor environment, \citet{rowe2021simple} proposed an airshed model that derives a similar infection probability formulation:
\begin{align}
    \mathbb{P}_{\text{Outdoor}}(p \rightarrow i \mid p \text{ infectious}) = 1 - \exp\left(\frac{-b\cdot q \cdot d_{i,p}}{Q^{\text{Outdoor}}} \right)
\end{align}
where $Q^{\text{Outdoor}} = L\cdot H\cdot V_{\infty}$ is the outdoor ventilation rate of clean air. $H$ and $L$ are the height and length (perpendicular to the wind direction) for a hypothetical outdoor modeling space. $V_{\infty}$ is the wind velocity ($m/h$). The suggested values for $H$ and $L$ are around 5$m$ and 50$m$, respectively. 

Hence, given the set of contact persons, the total infectious probability of $i$ during the segment $s$ due to close contact is 
\begin{align}
    \mathbb{P}_{\text{Cont}}^{(s)}(i \text{ infected}) = 1 - \prod_{p\in\mathcal{P}_s}\left[\big(1-\mathbb{P}_{\text{In/Outdoor}}(p \rightarrow i \mid p \text{ infectious})\big)\cdot\mathbb{P}(p \text{ infectious}) + \mathbb{P}(p \text{ not infectious}) \right]
    \label{eq_infect_s}
\end{align}
where Eq. \ref{eq_infect_s} is due to the fact that the probability of getting infected by at least one of $p \in \mathcal{P}_s$ equals one minus the probability of not getting infected by anyone.

\textbf{Infection by touching surfaces}: \citet{wilson2021modeling} estimate the infection probability of a single hand-to-fomite touch as a function of viral bioburden. Their estimation already accounts for uncertainties in transfer efficiency, fractions of the hand used for surface and face contact, and surface areas of the hand and of fomites available for contact. Define this probability as $\mathbb{P}_{\text{Touch}}(i\text{ infected}\mid V)$, where $V$ is the viral bioburden of this tough. For simplicity, let us assume the viral bioburden during a trip segment is a constant and the value is $V_s$. We also assume that the number of touches during segment $s$ (defined as $N_s$) is proportional to the duration of travel in $s$ (defined as $T_s$), and the factor is $\gamma$ (i.e., $N_s = \gamma_s \cdot T_s$). Therefore, the total infection probability for individual $i$ due to surface touching is:
\begin{align}
    \mathbb{P}_{\text{Surf}}^{(s)}(i \text{ infected}) = 1 - \big(1-\mathbb{P}_{\text{Touch}}(i\text{ infected}\mid V_s)\big)^{N_s}\label{eq_infect_tough}
\end{align}
The empirical values of $V_s$ can be obtained from \citet{harvey2020longitudinal} who collected viral bioburden data in daily activity environments. Since the infection risk for a single tough is small, Eq. \ref{eq_infect_tough} can be approximated by first-order Taylor series:
\begin{align}
    \mathbb{P}_{\text{Surf}}^{(s)}(i \text{ infected}) \approx \gamma_s \cdot T_s \cdot \mathbb{P}_{\text{Touch}}(i\text{ infected}\mid V_s) \label{eq_infect_tough_taylor}
\end{align}
where Eq. \ref{eq_infect_tough_taylor} is computationally more efficient than Eq. \ref{eq_infect_tough}.

\subsubsection{Infection risk by different travel modes}
Each segment $s\in\mathcal{S}_r$ is associated with a specific travel mode. Though we provide a general infection risk calculation model in Section \ref{sec_inf_risk_seg}, it is important to specify the model parameters and variables for each travel mode. In this section, we assume that the infectious environment is the same during the travel in a segment $s$ and the values are $b_s$, $q_s$, $Q_s^{\text{Indoor}}$, and $Q_s^{\text{Outdoor}}$. 

\textbf{Infection risk of transit segment}:
A transit segment (either bus or rail) usually includes multiple stops. Let the set of stops for the transit segment $s$ except for the last one be $\mathcal{A}_s$. We exclude the last stop because individual $i$ will alight when he/she arrives at the last stop. Let the set of passengers in a vehicle (exclude individual $i$) when the vehicle departs from station $a \in \mathcal{A}_s$ be $\mathcal{P}_{s, a}$. $\mathcal{P}_{s,a}$ can be obtained by smart card data and $\mathcal{P}_{s} = \cup_{a\in\mathcal{A}_s} \mathcal{P}_{s,a}$.  Let the vehicle travel time from station $a$ to the next stop be $TT_a$.

Then the infection probability for individual $i$ due to close contact in the vehicle when it travels from station $a$ to the next stop is:
\begin{align}
    \mathbb{P}_{\text{Cont}}^{(s,a)}(i \text{ infected})  = 1 - \prod_{p\in\mathcal{P}_{s,a}}\left[\mathbb{P}(p \text{ infectious})\cdot\exp{\left(\frac{-b_s\cdot q_s \cdot TT_a}{Q_s^{\text{Indoor}}} \right)} + (1 - \mathbb{P}(p \text{ infectious}))\right]\label{eq_inf_transit_a}
\end{align}
For any $p \in \mathcal{P}_{s,a}$, we can use smart card data to obtain their origin stations. Hence, $\mathbb{P}(p \text{ infectious})$ and $\mathbb{P}(p \text{ not infectious})$ can be approximated by regional infection statistics based on their origin stations. The total infection probability in the transit segment is:
\begin{align}
 \mathbb{P}_{\text{Cont }}^{(s)}(i \text{ infected})  = 1 - \prod_{a\in\mathcal{A}_{s}} \Big(1 - \mathbb{P}_{\text{Cont }}^{(s,a)}\Big) \quad \text{when $s$ is a transit segment} \label{eq_inf_transit_s}
\end{align}

Eqs. \ref{eq_inf_transit_a} and \ref{eq_inf_transit_s} may be computationally inefficient. When $\mathbb{P}(p \text{ infectious})$ is small, we can use the following approximation by ignoring all second-order multiplication terms with $\big(\mathbb{P}(p \text{ infectious})\big)^2$:
\begin{align}
&\prod_{p\in\mathcal{P}_{s,a}}\left[\mathbb{P}(p \text{ infectious})\cdot\exp{\left(\frac{-b_s\cdot q_s \cdot TT_a}{Q_s^{\text{Indoor}}} \right)} + (1 - \mathbb{P}(p \text{ infectious}))\right] \notag\\
&  = \prod_{p\in\mathcal{P}_{s,a}} \left[1 - \mathbb{P}(p \text{ infectious})\cdot\Bigg(1-\exp{\left(\frac{-b_s \cdot q_s \cdot TT_a}{Q_s^{\text{Indoor}}} \right) \Bigg)}  \right] \notag\\
& \approx 1 - \sum_{p \in {P}_{s,a}} \mathbb{P}(p \text{ infectious}) \cdot \Bigg(1 - \exp{\left(\frac{-b_s \cdot q_s \cdot TT_a}{Q_s^{\text{Indoor}}} \right)}\Bigg)
\end{align}
Then we have
\begin{align}
  \mathbb{P}_{\text{Cont}}^{(s,a)}(i \text{ infected})  \approx  \sum_{p \in \mathcal{P}_{s,a}} \mathbb{P}(p \text{ infectious}) \cdot \Bigg(1 - \exp{\left(\frac{-b_s\cdot q_s \cdot TT_a}{Q_s^{\text{Indoor}}} \right)}\Bigg) \label{eq_inf_transit_a_app}
\end{align}
which is simply the summation of probabilities of getting infected by anyone in $\mathcal{P}_{s,a}$. Similarly, if $\mathbb{P}_{\text{Cont }}^{(s,a)} $ is small, we can approximate $\mathbb{P}_{\text{Cont }}^{(s)}$ as:
\begin{align}
 \mathbb{P}_{\text{Cont}}^{(s)}(i \text{ infected})  \approx \sum_{a \in \mathcal{A}_{s}} \sum_{p \in \mathcal{P}_{s,a}} \mathbb{P}(p \text{ infectious}) \cdot \Bigg(1 - \exp{\left(\frac{-b_s\cdot q_s \cdot TT_a}{Q_s^{\text{Indoor}}} \right)}\Bigg)\notag\\  \text{when $s$ is a transit segment}\label{eq_inf_transit_s_app}
\end{align}
where Eqs. \ref{eq_inf_transit_a_app} and \ref{eq_inf_transit_s_app} are computationally more efficient because we replace the production with a summation. 

In terms of the surface-touching infection, we only need to specify $V_s$ (viral bioburden), $\gamma_s$ (touching rate), and $T_s$ (travel time) to use Eq. \ref{eq_infect_tough_taylor}. It is worth noting that $V_s$ can vary across different times of the day and transit routes according to the demand level. In general, times and routes with higher demand should have higher $V_s$. 



\textbf{Infection risk of walk/bike segment}: For a walk or bike segment, we assume there is no surface touching infection risk because the commuter does not need to tough public surfaces during commuting:
\begin{align}
    \mathbb{P}_{\text{Surf}}^{(s)}(i \text{ infected}) = 0 \quad \text{when $s$ is a bike/walk segment}
\end{align}
For the close-contact infection, we can approximate $\mathcal{P}_s$ from the pedestrian density data set. Denote the average contact time for an encounter as $d_{\text{W/B}}$. With the same approximation techniques, we can calculate the infection probability as  
\begin{align}
 \mathbb{P}_{\text{Cont}}^{(s)}(i \text{ infected})  \approx  |\mathcal{P}_{s}| \cdot \mathbb{P}(p \text{ infectious}) \cdot \Bigg(1 - \exp{\left(\frac{-b_s\cdot q_s \cdot d_{\text{W/B}}}{Q_s^{\text{Outdoor}}} \right)}\Bigg) \notag\\     \text{when $s$ is a bike/walk segment}\label{eq_inf_wb_s} 
\end{align}
where $\mathbb{P}(p \text{ infectious})$ can be obtained from the regional infectious statistics based on the walk/bike routes; $|\mathcal{P}_{s}|$ is the average number of encounters during the segment.

\textbf{Infection risk of ride-hailing segment}: When commuting with ride-hailing, $\mathcal{P}_s$ includes the driver and potential shared riders. $\mathcal{P}_s$ can be obtained from ride-hailing open source data to approximate the average number of persons in a vehicle. The contact time $d_{i,p}$ for the driver and individual $i$ is $T_s$ (total segment travel time). While $d_{i,p}$ for individual $i$ and shared riders will be approximated by shared rides data. With the same first-order approximation,  we can calculate the infection probability due to close contact as  
\begin{align}
 \mathbb{P}_{\text{Cont}}^{(s)}(i \text{ infected})  \approx \sum_{s\in \mathcal{P}_{s}} \mathbb{P}(p \text{ infectious}) \cdot \Bigg(1 - \exp{\left(\frac{-b_s\cdot q_s \cdot d_{i,p}}{Q_s^{\text{Indoor}}} \right)}\Bigg) \notag \\ \text{when $s$ is a ride-hailing segment}  \label{eq_inf_rh_s}  
\end{align}

In terms of surface-touching risks, it is possible that the previous riders are infectious and thus the seats are contaminated. We can also specify $V_s$ and $\gamma_s$ then use Eq. \ref{eq_infect_tough_taylor} for the infection risk calculation. However, it is worth noting that $V_s$ for ride-hailing should be much smaller than that of transit. 

\textbf{Infection risk of driving segment}: Driving is a private commute mode and there is generally no close contact or infectious surface touching during the travel. Hence, we simply assume $\mathbb{P}_{\text{Surf}}^{(s)}(i \text{ infected}) =\mathbb{P}_{\text{Cont }}^{(s)}(i \text{ infected}) = 0$ when $s$ is a driving segment. 

\subsection{Infection risk integration}
Section \ref{sec_inf_seg} provides the formulations for the calculation of infection risks for a specific segment. Since a commuting path $r$ consists of multiple segments, the total infection probability if using path $r \in \mathcal{R}_i$ is:
\begin{align}
    \mathbb{P}(i \text{ infected}\mid r) &= 1 - \prod_{s\in\mathcal{S}_r} \mathbb{P}(i \text{ not infected in segment $s$})\notag\\
    & = 1 - \prod_{s\in\mathcal{S}_r}  \left(1-\mathbb{P}_{\text{Surf}}^{(s)}(i \text{ infected})\right)\cdot\left(1-\mathbb{P}_{\text{Cont}}^{(s)}(i \text{ infected})\right)
\end{align}
Similarly, if the infectious probability is small in each segment, we have:
\begin{align}
\mathbb{P}(i \text{ infected}\mid r) \approx \sum_{s\in\mathcal{S}_r} \left( \mathbb{P}_{\text{Cont}}^{(s)}(i \text{ infected}) + \mathbb{P}_{\text{Surf}}^{(s)}(i \text{ infected}) \right)
\label{eq_inf_prob_path}
\end{align}
Combining with the path choice probabilities from Section \ref{sec_path}, we get the total infectious probability of individual $i$ during commuting:
\begin{align}
    \mathbb{P}(i \text{ infected}\mid o_i,d_i,t_i,m_i) = \sum_{r\in\mathcal{R}_i} \mathbb{P}_{\text{Path}}(r)\cdot \mathbb{P}(i \text{ infected}\mid r)
    \label{eq_inf_prob_final}
\end{align}

\subsection{Uncertainty quantification} \label{sec_uncertainty_quan}
Though Eq. \ref{eq_inf_prob_final} provides the final infection probability for a given trip, the estimation may suffer from errors due to uncertainties in parameters. In this section, we treat $\mathbb{P}(i \text{ infected}\mid o_i,d_i,t_i,m_i)$ as a random variable and aim to calculate its standard errors: $\sqrt{\text{Var}[{\mathbb{P}}(i \text{ infected}\mid o_i,d_i,t_i,m_i)]}$.

In this study, we focus on the uncertainty due to the infection calculations. Hence, let us assume $\mathbb{P}_{\text{Path}}(r)$ is fixed. Then:
\begin{align}
    \text{Var}[{\mathbb{P}}(i \text{ infected}\mid o_i,d_i,t_i,m_i)] = \sum_{r\in\mathcal{R}_i} \big(\mathbb{P}_{\text{Path}}(r)\big)^2\cdot\text{Var}[\mathbb{P}(i \text{ infected}\mid r)]
    \label{eq_var_p1}
\end{align}

From Eq. \ref{eq_inf_prob_path}, we can further decompose the variance to different segments due to the independence across segments:
\begin{align}
    \text{Var}[\mathbb{P}(i \text{ infected}\mid r)] = \sum_{s\in\mathcal{S}_r} \left( \text{Var}\big[\mathbb{P}_{\text{Cont}}^{(s)}(i \text{ infected})\big] + \text{Var}\big[\mathbb{P}_{\text{Surf}}^{(s)}(i \text{ infected})\big] \right)
\end{align}

Therefore, we only need to specify $\text{Var}\big[\mathbb{P}_{\text{Cont}}^{(s)}(i \text{ infected})\big]$ and $\text{Var}\big[\mathbb{P}_{\text{Surf}}^{(s)}(i \text{ infected})\big]$ for different types (i.e., travel modes) of segments. 

\textbf{Transit segment}: According to Eq. \ref{eq_inf_transit_s_app}, the infection probability due to close contact in public transit is simply the probability summation over different stations and encounters. In the real world, $\mathcal{P}_{s,a}$ is uncertain due to demand variations in public transit systems. However, it is generally hard to model the uncertainty of a set. Therefore, let us define the average probability that a passenger in $\mathcal{P}_{s,a}$ is infectious as $\Bar{\mu}_{s,a}$. Then Eq. \ref{eq_inf_transit_s_app} can be rewrite as:
\begin{align}
     \mathbb{P}_{\text{Cont}}^{(s)}(i \text{ infected})  \approx \sum_{a \in \mathcal{A}_{s}} |\mathcal{P}_{s,a}| \cdot \Bar{\mu}_{s,a} \cdot \Bigg(1 - \exp{\left(\frac{-b_s\cdot q_s \cdot TT_a}{Q_s^{\text{Indoor}}} \right)}\Bigg)  \quad \text{when $s$ is a transit segment}
\end{align}
where $|\mathcal{P}_{s,a}|$ is the number of passengers in the vehicle (excluding $i$) when the vehicle departs from station $a$. 
The variance can be formulated as:
\begin{align}
     \text{Var}[\mathbb{P}_{\text{Cont}}^{(s)}(i \text{ infected})] = \sum_{a \in \mathcal{A}_{s}} \text{Var}\left[|\mathcal{P}_{s,a}| \cdot \Bar{\mu}_{s,a} \cdot \Bigg(1 - \exp{\left(\frac{-b_s\cdot q_s \cdot TT_a}{Q_s^{\text{Indoor}}} \right)}\Bigg)\right]  \notag \\ \text{when $s$ is a transit segment}
     \label{eq_var_transit}
\end{align}

Deriving the closed-form formulation for the variance of the product of several random variables (or the exponential of several variables) is difficult. In some simple cases, one may use Jensen's inequality to get the variance lower (or upper) bound. However, consider a general function $f(\boldsymbol{Z})$ (may not be convex or concave), where $\boldsymbol{Z}$ is a vector of random variables. Obtaining the analytical form of $\text{Var}[f(\boldsymbol{Z})]$ is generally hard. Therefore, we propose a bootstrapping-based algorithm to estimate the empirical variance (Algorithm \ref{alg_bootstr}). The inputs of the algorithm are $f(\boldsymbol{Z})$,  the distribution of $\boldsymbol{Z}$ (i.e., $\mathbb{P}(\boldsymbol{Z})$), and maximum sample times $M$. The idea is to sample $\boldsymbol{Z} \sim \mathbb{P}(\boldsymbol{Z})$ and use samples to estimate the variance. 

\begin{algorithm}[htb]
\caption{Bootstrapping-based empirical variance estimation algorithm}
\label{alg_bootstr}
\begin{algorithmic}[1]
\Function{Bootstrapping-Variance}{$f(\boldsymbol{Z})$, $\mathbb{P}(\boldsymbol{Z})$, $M$}
    \For {$m = 1,2,...,M$}
        \State Sample $\boldsymbol{Z}^{(m)} \sim \mathbb{P}(\boldsymbol{Z})$
    \EndFor
    \State $\Bar{f}(\boldsymbol{Z})= \frac{1}{M}\sum_{m=1}^M {f}(\boldsymbol{Z}^{(m)})$ \Comment{Estimate empirical mean}
    \State $\text{Var}[f(\boldsymbol{Z})] = \frac{1}{M}\sum_{m=1}^M\big({f}(\boldsymbol{Z}^{(m)}) - \Bar{f}(\boldsymbol{Z})\big)^2$ \Comment{Estimate empirical variance}
    \State \textbf{return} $\text{Var}[f(\boldsymbol{Z})]$
\EndFunction
\end{algorithmic}
\end{algorithm}
It is worth noting that, the sampling process in Algorithm \ref{alg_bootstr} (Line 3) can be done independently, jointly, or sequentially, depending on whether the elements in $\boldsymbol{Z}$ are independent, dependent, or conditionally independent. 

Given Algorithm \ref{alg_bootstr}, we can estimate $\text{Var}\left[|\mathcal{P}_{s,a}| \cdot \Bar{\mu}_{s,a} \cdot \big(1 - \exp{\left(\frac{-b_s\cdot q_s \cdot TT_a}{Q_s^{\text{Indoor}}} \right)}\big)\right]$ through the distribution of $|\mathcal{P}_{s,a}|$,  $\Bar{\mu}_{s,a}$, $b_s$, $q_s$, $TT_a$, and $Q_s$. In this study, we assume that $|\mathcal{P}_{s,a}|$ (vehicle load) and $TT_a$ (travel time) are normally distributed based on the observations in the empirical data. Their distribution parameters can be estimated from the smart card and automated vehicle location data. $b_s$, $q_s$, and $Q_s$ are uniformly distributed and their distributions are shown in Table \ref{tab_inf_para} based on the literature. For $\Bar{\mu}_{s,a}$, the distribution is hard to parameterize because it is generated by sampling different $\mathcal{P}_{s,a}$. We, therefore, keep all samples for $\Bar{\mu}_{s,a}$ as an empirical distribution. Then every sample of $\Bar{\mu}_{s,a}\sim\mathbb{P}(\Bar{\mu}_{s,a})$ is essentially a bootstrap from its sample pool. 

In terms of surface touching infection, from Eq. \ref{eq_infect_tough_taylor}, we use Algorithm \ref{alg_bootstr} to estimate the variance by setting $f(\boldsymbol{Z}) = \gamma_s \cdot T_s \cdot \mathbb{P}_{\text{Touch}}(i\text{ infected}\mid V_s)$ and $\boldsymbol{Z} = (\gamma_s, T_s, V_s, \mathbb{P}_{\text{Touch}}(i\text{ infected}\mid V_s)$. The distribution of $ \mathbb{P}_{\text{Touch}}(i\text{ infected}\mid V_s)$ can be obtained from the data in \citet{wilson2021modeling}. The distribution of $V_s$ can be obtained from the results in \citet{harvey2020longitudinal}. The distribution of $T_s$ can be obtained from the AVL data. $\gamma_s $ is assumed to be uniformly distributed and its distribution is given in Table \ref{tab_inf_para}.

\textbf{Walk/Bike segment}:
The variance of contact-based infection risks in the walk and bike segments, according to Eq. \ref{eq_inf_wb_s}, can be expressed as:
\begin{align}
 \text{Var}[\mathbb{P}_{\text{Cont}}^{(s)}(i \text{ infected})] = \text{Var}\left[|\mathcal{P}_{s}| \cdot  \Bar{\mu}_s \cdot \Bigg(1- \exp{\left(\frac{-b_s\cdot q_s \cdot d_{\text{W/B}}}{Q_s^{\text{Outdoor}}} \right)\Bigg)}\right] \notag\\     \text{when $s$ is a bike/walk segment}\label{eq_inf_wb_s_var} 
\end{align}
where $\Bar{\mu}_s$ is the average infectious probability of encounters in $\mathcal{P}_{s}$. It has a similar formulation as Eq. \ref{eq_var_transit}, thus can be calculated using Algorithm \ref{alg_bootstr}. The distribution of $|\mathcal{P}_{s}|$ can be obtained from pedestrian flow data. The distribution of $\Bar{\mu}_s$ can be obtained from regional statistics. All infection-related parameters (including $d_{\text{W/B}}$) are assumed to be uniformly distributed and their distributions are shown in Table \ref{tab_inf_para}. Since we assume there are no infection risks related to surface touching for the walk and bike segments, we do not need to consider their variances.

\textbf{Ride hailing segment}:
For the ride-hailing segment, since $\mathcal{P}_s$ is relatively small (maximum three riders in a vehicle), we assume $|\mathcal{P}_s|$ follows a multinomial distribution across $\{0,1,2,3\}$ (i.e., maximum 2 other passengers plus 1 driver). The specific distribution can be obtained from ride-sharing data. To get $\text{Var}[\mathbb{P}_{\text{Cont}}^{(s)}(i \text{ infected})]$ for ride-hailing segments, the sampling process is as follows (slightly different from Algorithm \ref{alg_bootstr}):
\begin{itemize}
    \item Step 1: Sample $|\mathcal{P}_s|$ from the multinomial distribution. 
    \item Step 2: Based on the value of $|\mathcal{P}_s|$, sample the same number of passengers and drivers to generate the $|\mathcal{P}_s$, for each passenger $p \in \mathcal{P}_s$, we also sample $d_{i,p}$ (shared trip time). 
    \item Step 3: Sample other infection-related parameters from the distribution defined in Table \ref{tab_inf_para}. Calculate the infection probability close contact based on Eq. \ref{eq_inf_rh_s}.
\end{itemize}
After $M$ samples, we can estimate the sample variance as an approximation of $\text{Var}[\mathbb{P}_{\text{Cont}}^{(s)}(i \text{ infected})]$ similar to Algorithm \ref{alg_bootstr}. The variance of surface touching-based infection probability is calculated in the same way as a transit segment except for replacing the distributions of $T_s$, $V_s$, and $\gamma_s$.  

For the driving segment, since we assume there is no infection risk, the variances are not calculated.  

\section{Case study}\label{sec_case_study}
The case study is based on available data from MIT. The model is implemented for the commuting of all MIT students and staff in the greater Boston area.
\subsection{Data sources and parameter settings}
We use data from various sources in this study to estimate the infection risk and set up parameters. The first data set is the MIT staff commuting survey. The survey collects ($o_i$, $d_i$, $t_i$, $m_i$) of every individual $i \in \mathcal{I}$. Given this information, for every individual $i$, we generate the set of paths $\mathcal{R}_i$ and the associated path attributes (i.e., walking time, waiting time, in-vehicle time, travel cost) based on the Google map API. For simplicity, we only consider the first path associated with the travel mode $m_i$ recommended by google map API (i.e., ignoring the path choice estimation). This assumption is reasonable because in most cases there is only one transit route available if $m_i= \text{Transit}$. For driving, biking, or ride-hailing, individuals usually follow the navigation system and thus choose the first option provided by the google map API. 

Another data source we use for infection risk calculation in the transit segment is the smart card data from the Massachusetts Bay Transportation Authority (MBTA). We use the historical smart card data at the same time of the day (one-hour interval) in the last 2 weeks to calculate the mean and variance of the number of passengers in $\mathcal{P}_{s,a}$. Specifically, the smart card data provides the tap-in time and locations. We adopted the destination estimation model proposed by \citet{sanchez2017inference} to obtain the destination of each trip. Then, a network loading model \citep{mo2020capacity} is used to obtain the vehicle load at each time interval. The mean infectious probability $\bar{\mu}_{s,a}$ for passengers in the vehicle is calculated based on regional statistics and their origins. The travel time between stops (i.e., $TT_a$) is obtained from automated vehicle location (AVL) data. 

In terms of the bike and walk segments, there is no open-source pedestrian and cyclist density data available for this study. Therefore, we generate the synthetic pedestrian and cyclists density data by combining population data in Boston \citep{mass2020population}, Massachusetts Travel Survey (MTS) \citep{boston2011mpo}, and national household travel survey (NHTS) data \citep{nhts2017}. The synthetic pedestrian density data include the mean and variance of the number of cyclists and pedestrians at a specific street for each time interval (in this study, every 1 minute). The detailed data generation process is shown in \ref{app_pedestrain_data}. The basic idea is to generate many bike and walk trajectory samples using the available dataset and aggregate these samples at the street-minute level. Therefore, given a bike/walk segment $s$ of individual $i$, based on its trajectory, we can sample the number of bike/walk encounters using the generated cyclists and pedestrians density data above. This process is replicated multiple times to get the distribution of $|\mathcal{P}_s|$. The distribution of $\bar{\mu}_s$ is calculated based on neighborhood infection rates \citep{adam2020coviddata}. The contact time ($d_{\text{W/B}}$) is assumed to be uniformly distributed with $\mathcal{U}(4,6)$ seconds for a walking trip and $\mathcal{U}(2,4)$ seconds for a biking trip. 

For the ride-hailing segment, we only need the distribution of $|\mathcal{P}_s|$ (number of passengers), $d_{i,p}$ (shared trip duration). However, there is no public ride-hailing data for Boston. In this study, we use the open-source Transportation Network Company (TNC) data in Chicago \citep{chicago2018tncdata} to calculate the distribution of $|\mathcal{P}_s|$ and $d_{i,p}$ as an approximation for those of Boston.

Through the Google Map API, we can obtain the value of $T_s$. However, the distribution of $T_s$ is unknown. Ideally, the distribution of $T_s$ can be obtained from GPS data. Given the data limitations, we assume $T_s$ is normally distributed and the value obtained from the Google Map API is the mean. The standard deviation is assumed to be $T_s\times 30\%$ based on the empirical study \citep{li2013empirical}. It is worth noting that, for the transit segment, instead of using Google Map data, we obtain the stop-to-stop travel time using AVL data. Hence, we calculate the segment travel time as $T_s =\sum_{a\in\mathcal{A}_s} TT_{a}$. The mean and variance of $T_s$ are just the summations of the mean and variance of $TT_{a}$, respectively, by assuming independence across segments. 


The infection-related parameters are summarized in Table \ref{tab_inf_para}:
\begin{table}[htbp]
\centering
\caption{Infection-related parameters}\label{tab_inf_para}
\resizebox{\textwidth}{!}
{
{\renewcommand{\arraystretch}{1} 
\begin{tabular}{@{}ccccccccc@{}}
\toprule
\multirow{2}{*}{Mode} & \multicolumn{8}{c}{Parameters  and distribution}                                                                                                                                                                                  \\ \cmidrule(l){2-9} 
                      & \multicolumn{1}{c}{$b_s$ ($m^3/h$)} & \multicolumn{1}{c}{$q_s$ ($/h$)} & \multicolumn{1}{c}{$V_{\infty}$ ($m/s$)} & \multicolumn{1}{c}{$L$ ($m$)} & \multicolumn{1}{c}{$H$ ($m$)} & \multicolumn{1}{c}{$Q_s^{\text{Indoor}}$ ($m^3/h$)} & \multicolumn{1}{c}{$\gamma_s$ ($/h$)} & \multicolumn{1}{c}{$V_s$ (gc/$cm^2$) }\\ \midrule
Transit (Train)              & $\mathcal{U}(0.65, 0.79)$  & $\mathcal{U}(1, 31)$  & N.A. &  N.A.    &  N.A.    &    $\mathcal{U}(800, 1100)$      &  $\mathcal{U}(3, 5)$  & $\mathcal{U}(30, 100)$ \\
Transit (Bus)              & $\mathcal{U}(0.65, 0.79)$  & $\mathcal{U}(1, 31)$  & N.A. &  N.A.    &  N.A.    &    $\mathcal{U}(300, 500)$      &  $\mathcal{U}(3, 5)$  & $\mathcal{U}(30, 100)$ \\
Bike                  & $\mathcal{U}(1.4, 1.8)$          &  $\mathcal{U}(2, 100)$      & $\mathcal{U}(2, 4)$    & $\mathcal{U}(30, 60)$               &  $\mathcal{U}(2.5, 5)$                    &   N.A.                          &    N.A.    &    N.A.                    \\
Walk                  & $\mathcal{U}(1.2, 1.6)$      &   $\mathcal{U}(2, 100)$       &   $\mathcal{U}(1, 2)$    &    $\mathcal{U}(30, 60)$        &          $\mathcal{U}(2.5, 5)$             &     N.A.                        &  N.A.                        &  N.A.     \\
Ride-hailing          & $\mathcal{U}(0.65, 0.79)$    &   $\mathcal{U}(1, 31)$    &  N.A.                          &   N.A.    &    N.A.         &     $\mathcal{U}(90, 120)$                      &    $\mathcal{U}(1, 3)$       &   $\mathcal{U}(2, 50)$  \\ \bottomrule

\multicolumn{9}{l}{\begin{tabular}[c]{@{}l@{}} N.A.: Not applicable\\
The parameters for transit infections are obtained from \citet{zhou2021virus}. \\
The parameters for quanta generation rates are obtained from \citet{buonanno2020estimation} and \citet{buonanno2020quantitative}. \\
The parameters for outdoor infection models are obtained from \citet{rowe2021simple}.\\
The ride-hailing ventilation rate is based on \citet{ott2008air}.\\
The viral bioburden data is adapted from \citet{harvey2020longitudinal} (assuming 10\% viruses are infectious).\\
\end{tabular}}
\end{tabular}
}
}
\end{table}

\subsection{Data description and statistics}
\begin{figure}[htb]
\centering
\subfloat[Departure time distribution]{\includegraphics[width=0.33\linewidth]{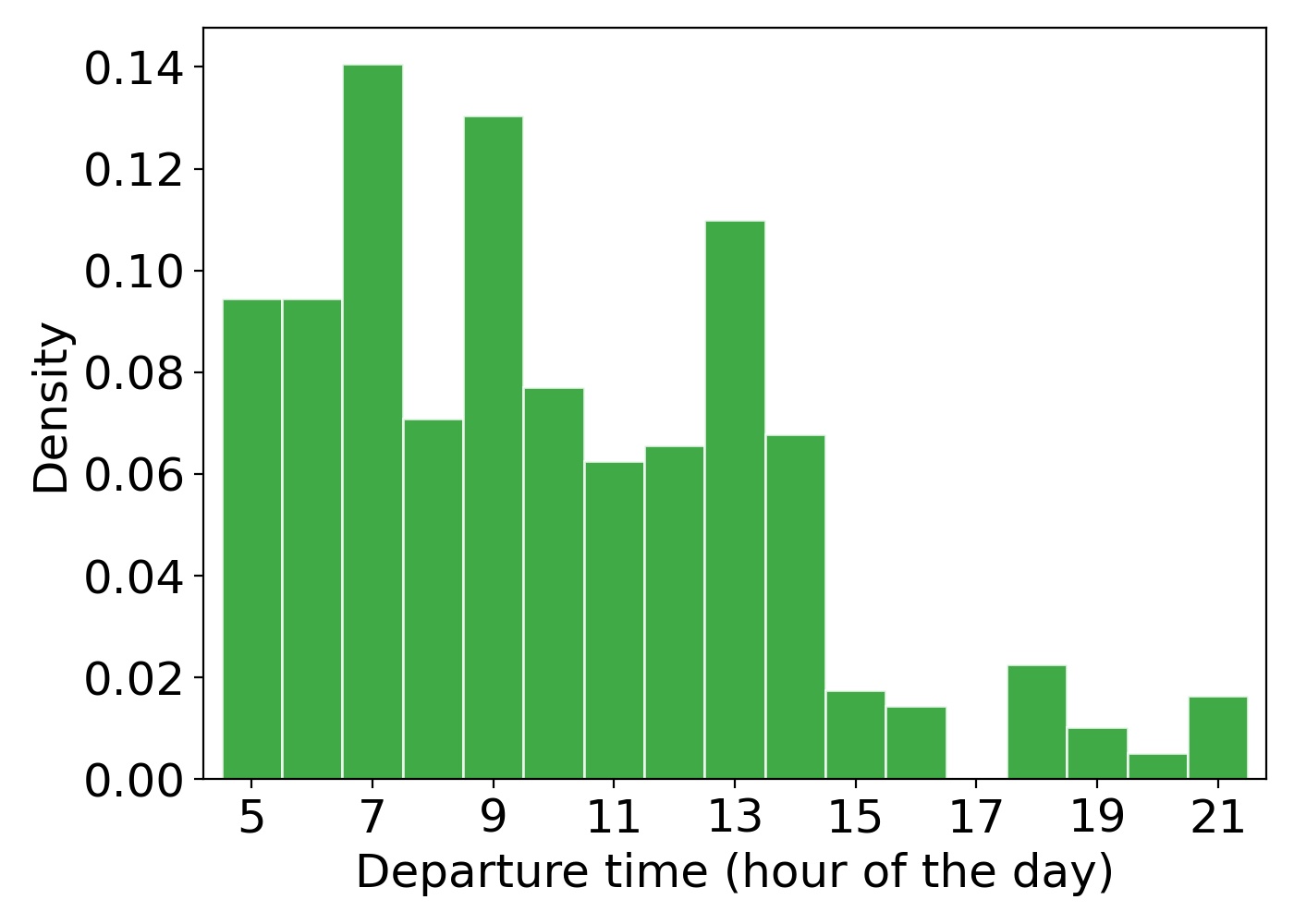}\label{fig_departure}}
\hfil
\subfloat[Travel mode distribution]{\includegraphics[width=0.33\linewidth]{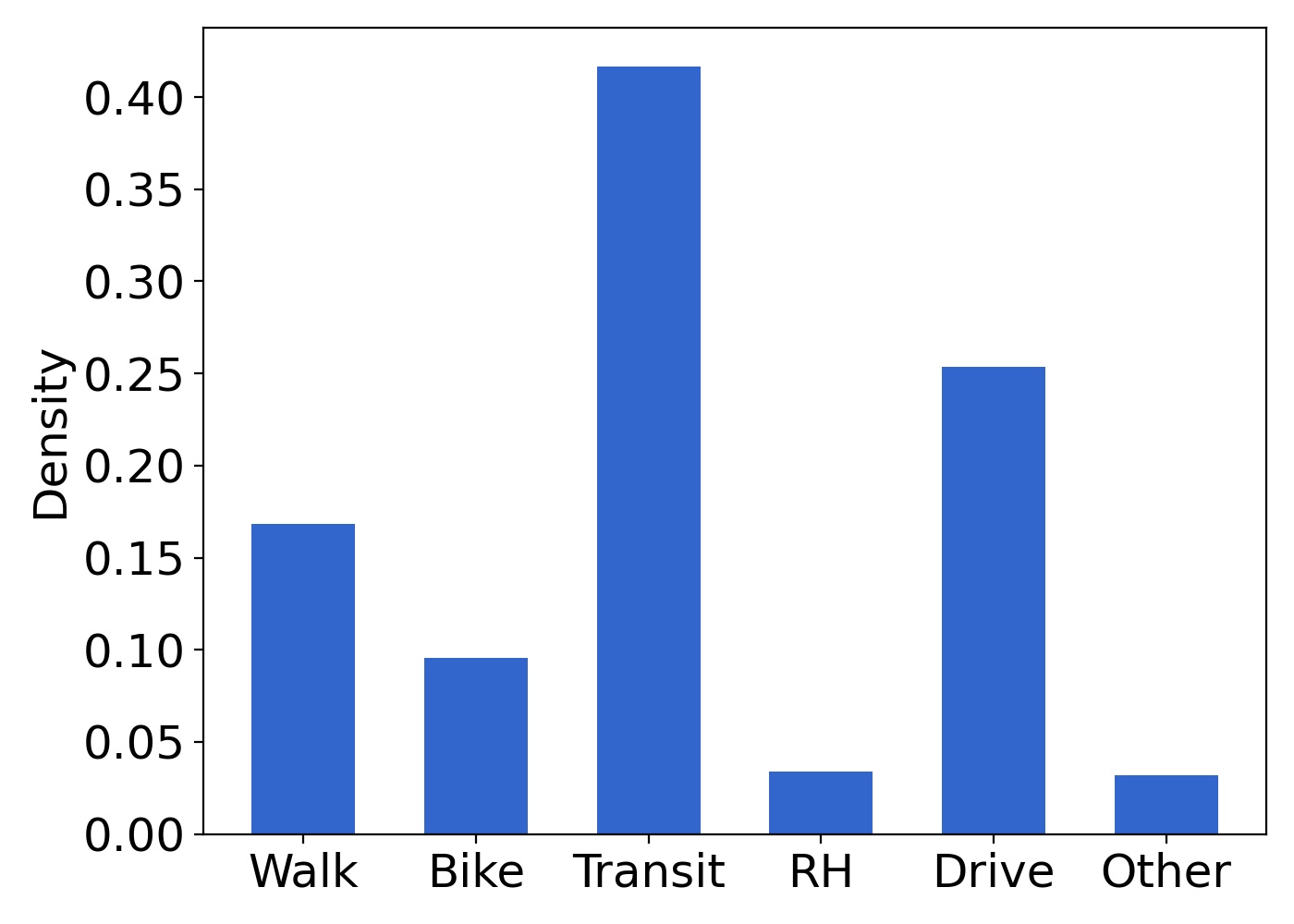}\label{fig_mode}}
\hfil
\subfloat[Person type distribution]{\includegraphics[width=0.33\linewidth]{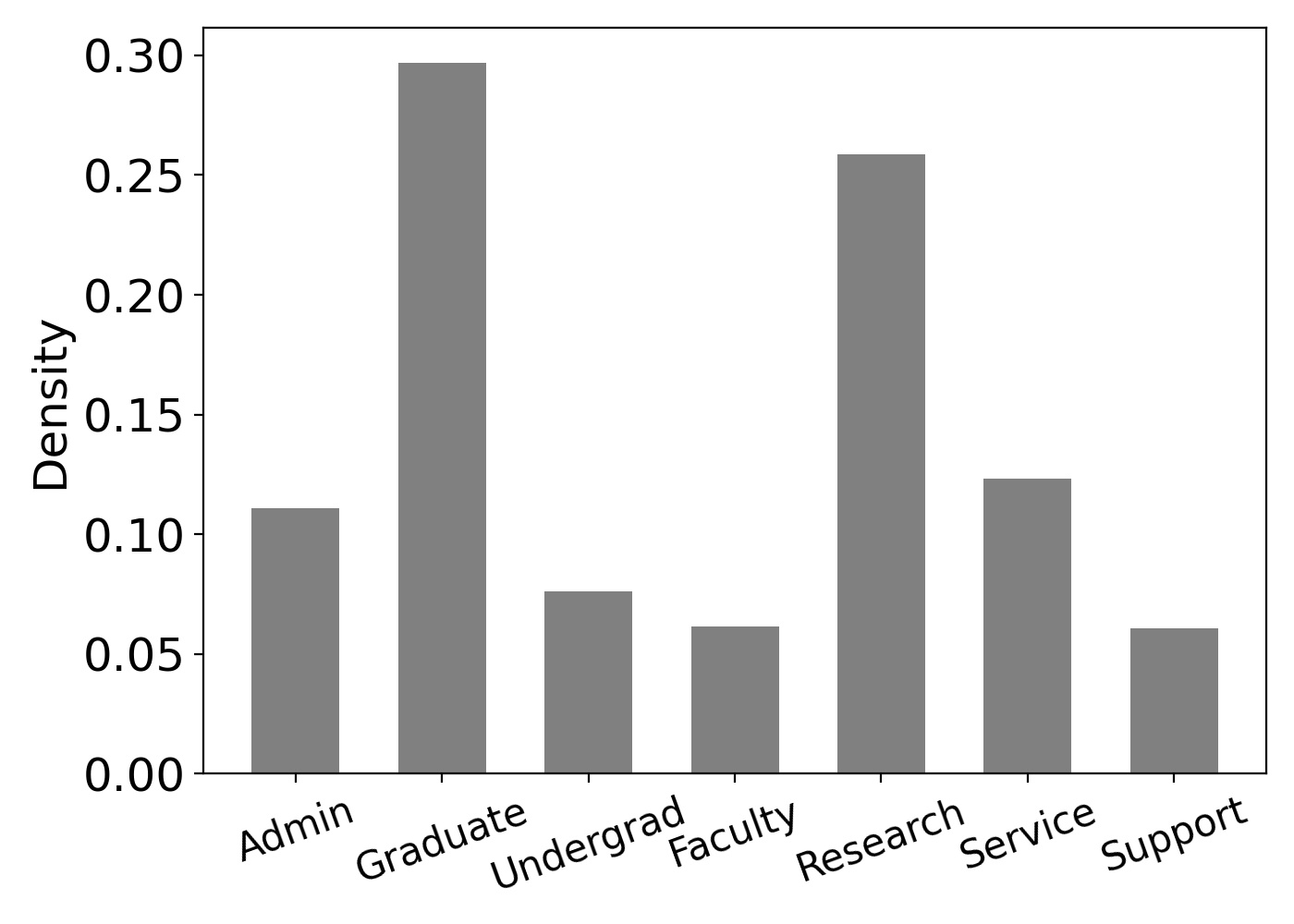}\label{fig_job}}
\caption{Descriptive statistics of MIT staff commuting survey (RH indicates Ride-hailing)}
\label{fig_mode_dep}
\end{figure}

The MIT staff commuting survey consists of 974 individual responses with information on $(o_i,d_i,t_i,m_i)$. Note that only commuting trips toward MIT are considered. The distributions of their departure times, travel modes, and job categories are shown in Figure \ref{fig_mode_dep}.

Since only trips toward MIT are considered, most of their departure times are in the morning hours (Figure \ref{fig_departure}). There are also some people going to campus in the evening, which may be students living close by or staff on night shifts. In terms of travel modes (Figure \ref{fig_mode}), more than 40\% of the MIT staff and students choose transit as their commuting mode. Driving is the second most popular mode. Of all individuals filling out the survey, around 30\% are graduate students. Respondents also include faculty and staff (e.g., admin, service, support, research). 

\begin{figure}[htb]
\centering
\subfloat[Number of segments distribution]{\includegraphics[width=0.45\linewidth]{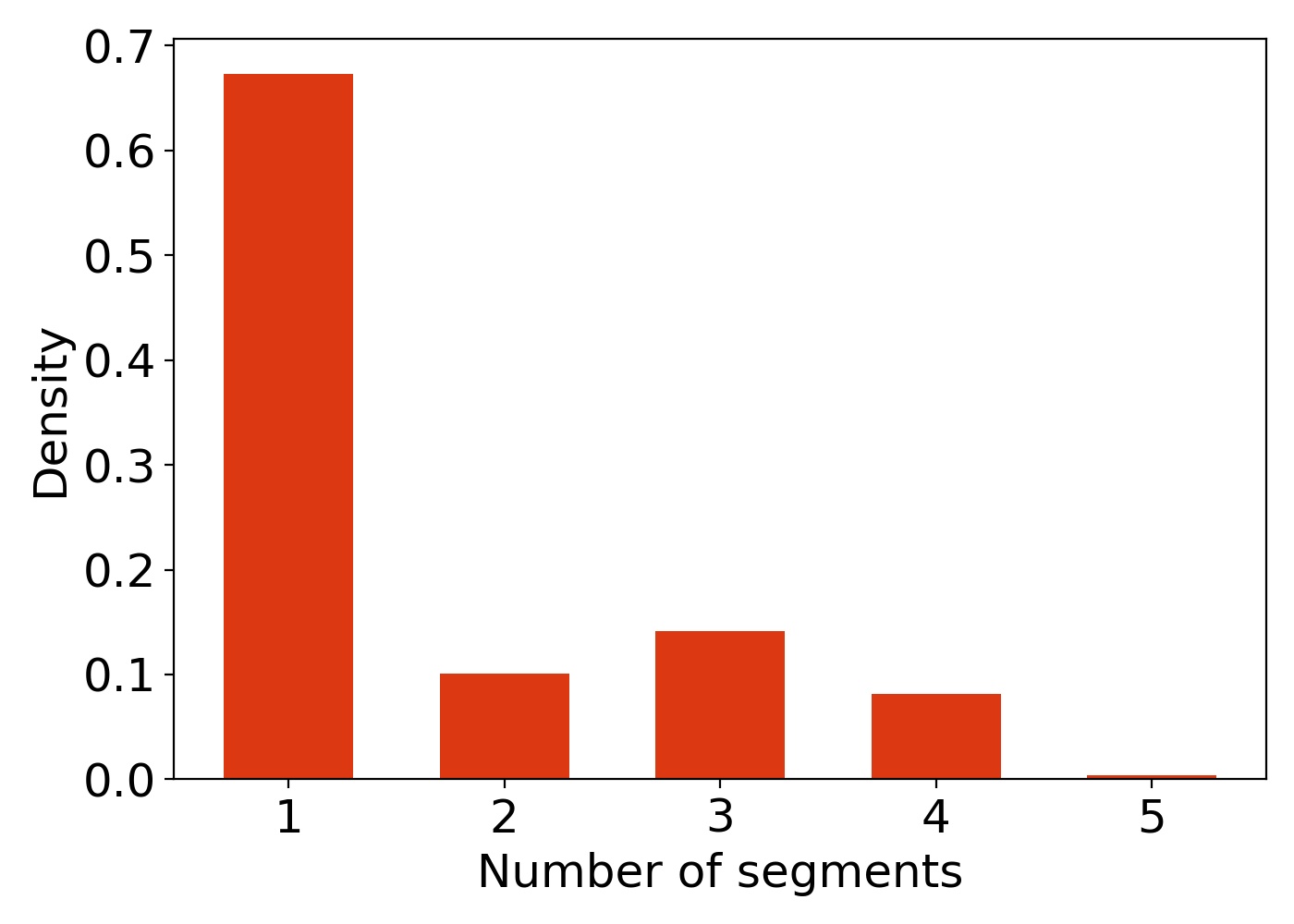}\label{fig_num_seg}}
\hfil
\subfloat[Travel mode distribution]{\includegraphics[width=0.45\linewidth]{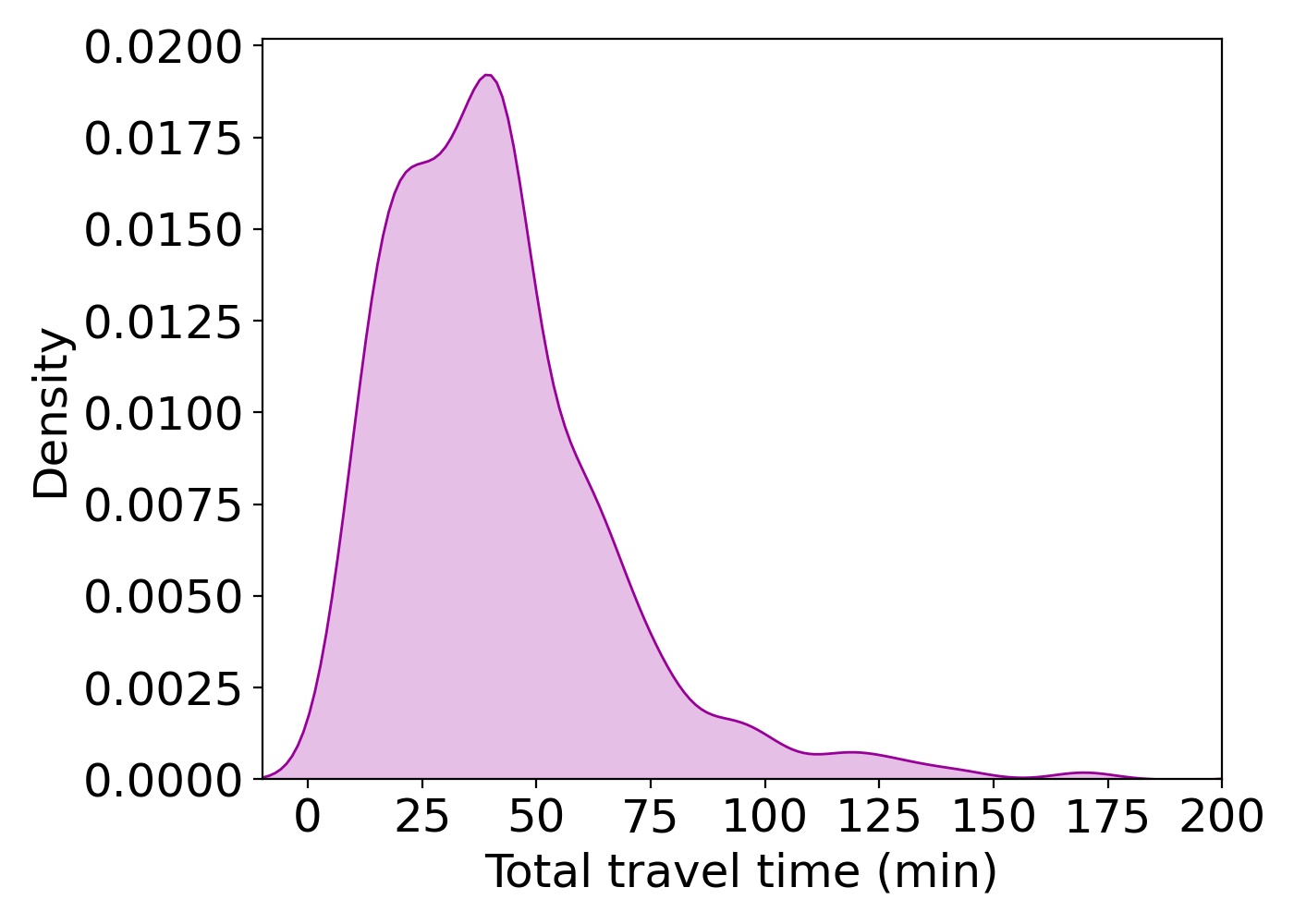}\label{fig_tt}}
\caption{Distribution of the number of segments and travel times}
\label{fig_tt_num_seg}
\end{figure}

Figure \ref{fig_tt_num_seg} shows the distribution of travel information collected by the Google Map API. Most of the commuting trips have only one segment (Figure \ref{fig_num_seg}). The maximum number of segments is five. The travel times of all trips are approximately log-normal distributed with a long tail (Figure \ref{fig_tt}). The travel times for most of the trips are within 1 hour. 

\begin{figure}[htb]
\centering
\subfloat[Travel time to next stop]{\includegraphics[width=0.45\linewidth]{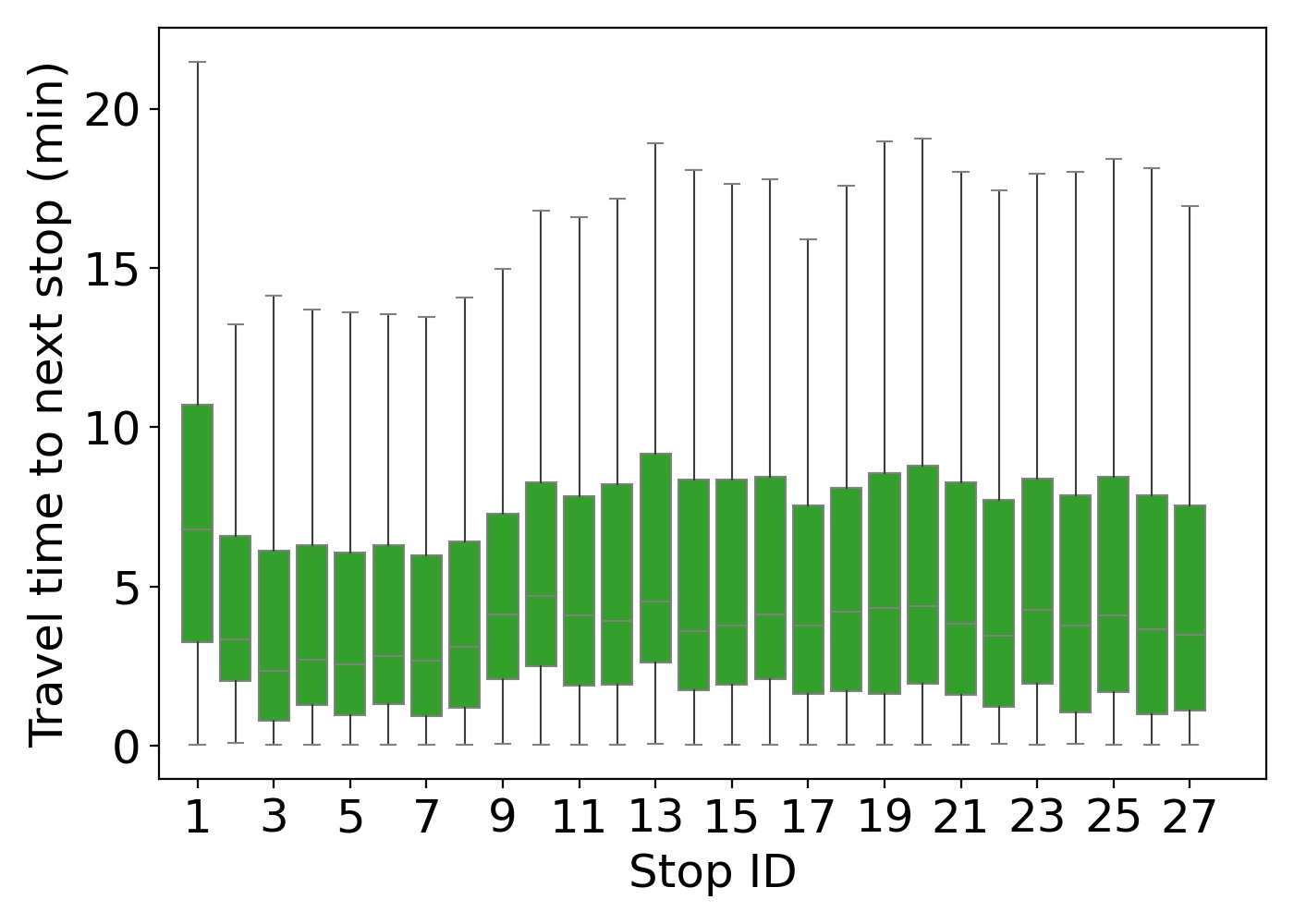}\label{fig_tt_bus}}
\hfil
\subfloat[Passenger load]{\includegraphics[width=0.45\linewidth]{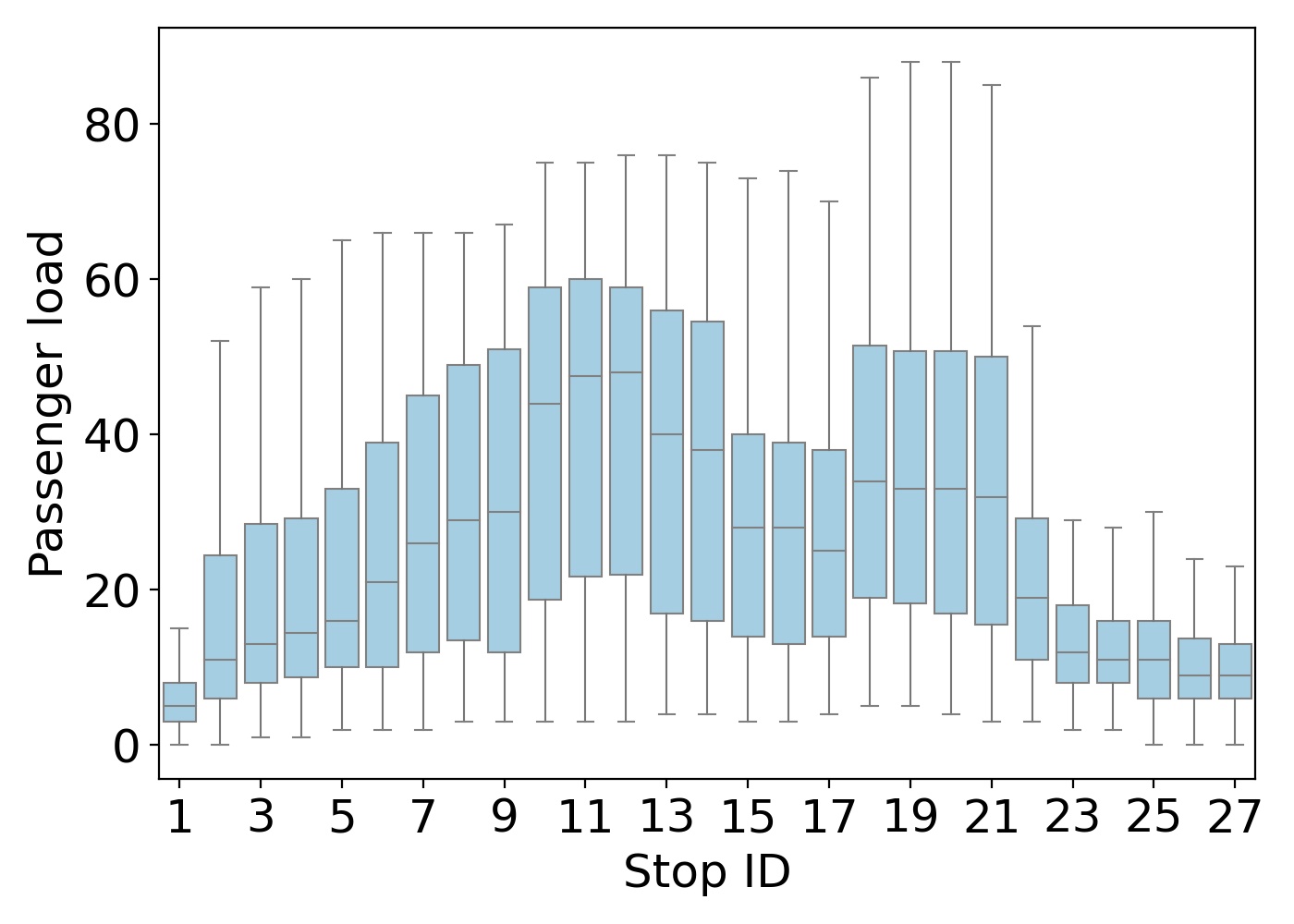}\label{fig_pass}}
\caption{Box plots of travel time and passenger load for Bus Route 1 (Stop ID represents the stop sequences from north to south)}
\label{fig_load_tt}
\end{figure}

For transit trips, Figure \ref{fig_load_tt} shows the box plots of travel times between two consecutive stations and the associated passenger load for Bus Route 1 in Boston. Route 1 is a popular bus service along Massachusetts Avenue. The travel time to the first stop is relatively large as vehicles usually depart from garages (Figure \ref{fig_tt_bus}). In terms of passenger load (Figure \ref{fig_pass}), the middle stops in the route have a relatively larger number of passengers onboard. The graphs show that both travel time and passenger loads are uncertain and the uncertainties should be captured in the virus transmission modeling. 

\begin{figure}[htb]
\centering
\subfloat[Number of bike trips]{\includegraphics[width=0.48\linewidth]{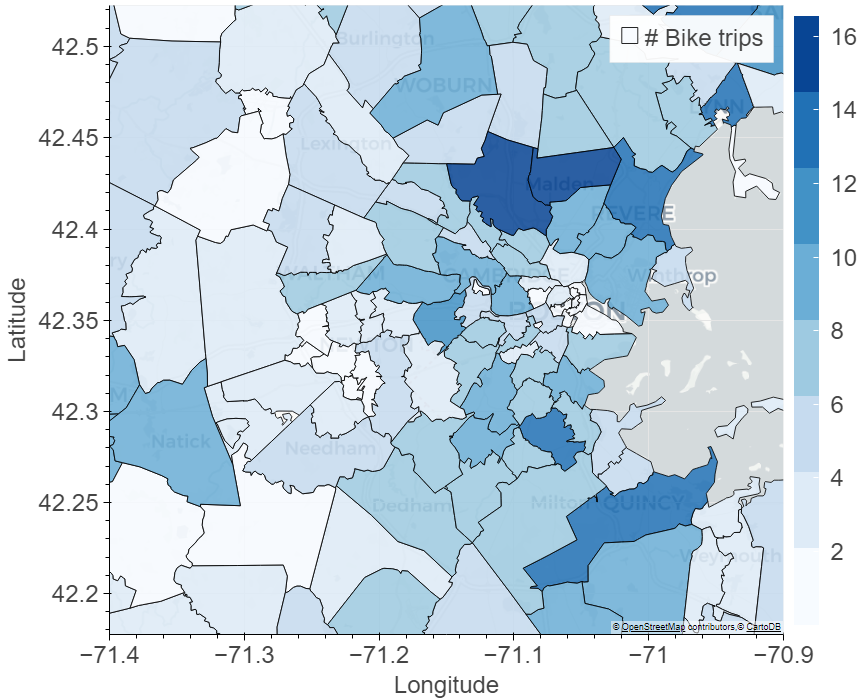}\label{fig_num_bike}}
\hfil
\subfloat[Number of walk trips]{\includegraphics[width=0.48\linewidth]{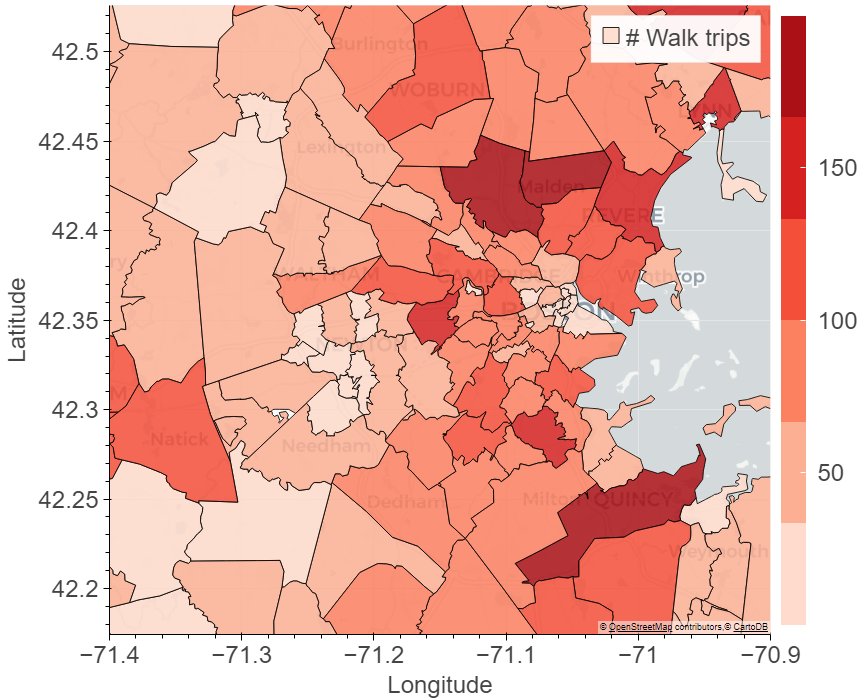}\label{fig_num_walk}}
\caption{Spatial distribution for the number of walk and bike trips at 8:00 AM}
\label{fig_bike_walk}
\end{figure}

Figure \ref{fig_bike_walk} shows the spatial distribution for the inferred number of walk and bike trips at 8:00 AM (details in \ref{app_pedestrain_data}). Their spatial distributions are similar. More trips happen at places with higher population densities. The number of walking trips is larger than that of bike trips. 

\subsection{Results}
\subsubsection{Infection risk for individuals}
\begin{figure}[htb]
\centering
\includegraphics[width = 0.8\linewidth]{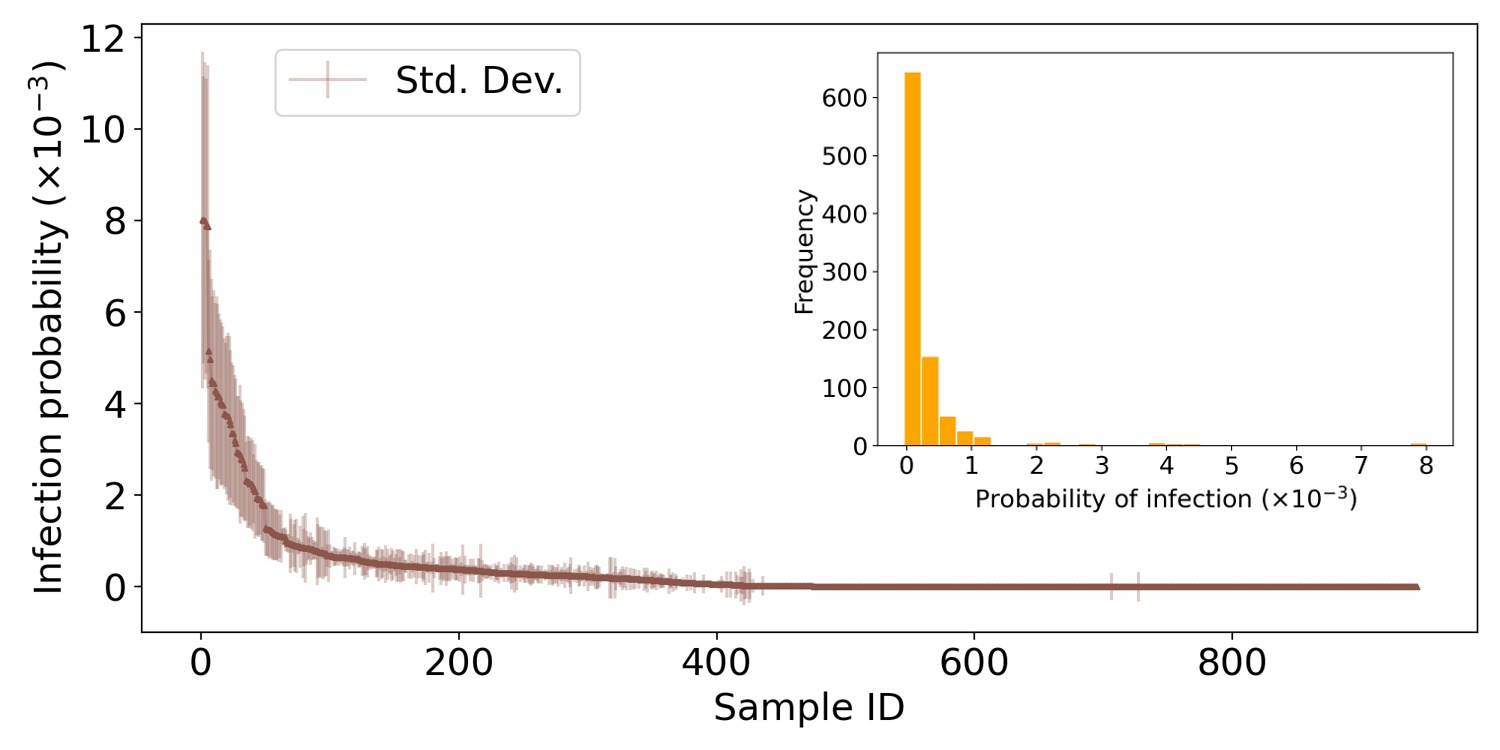}
\caption{Inferred individual infection probabilities (sorted by values in descending order)}
\label{fig_individual}
\end{figure}

After excluding individuals with ``other'' travel modes. We have 943 remaining individuals. We calculate their infection probabilities and standard deviations using the proposed method. Results are shown in Figure \ref{fig_individual}. Most of the individuals have an infection probability close to zero. The maximum infection probability is around 0.8\%. This implies that the probability of getting infected during commuting to MIT is quite low. The results are consistent with the previous studies modeling infection risks in transit \citep{zhou2021virus}. In terms of the estimation errors, the standard deviations are approximately 50\% of the estimated probability.  

Since individuals with different travel modes, departure times, and travel distances may have different infection probabilities. We run a linear regression model to analyze the impact of different factors on infection risks. The dependent variable is the inferred infection probability and the independent variables are as follows:
\begin{itemize}
    \item If faculty: Whether the individual is a faculty at MIT (Yes = 1).
    \item Distance (km): Euclidean distance from the individual's home to MIT.  
    \item If transit: Whether the individual's commuting mode is transit (Yes = 1).
    \item If morning peak: Whether the individual's departure time is between 6:00 AM and 9:00 AM (Yes = 1).
\end{itemize}

The results of the linear regression are shown in Table \ref{tab_factor_pred}. ``Distance$\times$If transit'' is added to differentiate the distance impact for transit and non-transit users. We find that the faculty is less likely to get infected. The reason may be that they usually drive to school and driving is the safest travel mode in terms of infection protection. Individuals who live further from the school and commute by transit have a higher infection risk. This may be due to the fact that they have a longer travel time and more close contacts, thus are more likely to be exposed to viruses. We also observe people with a departure time in the morning peak have a higher probability of being infected, which may be due to the larger amount of encounters in the morning peak hours.  It is worth noting that ``Distance'' and ``If transit'' are not significant unless combining them together (i.e., ``Distance$\times$If transit''). This implies that the impact of distance on infection risks mainly applies to transit users.

\begin{table}[htbp]
\centering
\caption{Factors on infection probability (\%)}\label{tab_factor_pred}
\begin{tabular}{@{}lclc@{}}
\toprule
Variable  & Coefficient (p-value)    & Variable       & Coefficients (p-value)   \\ \midrule
Intercept  & -0.015 (0.997) & Distance   & 0.048 (0.834) \\
If transit & 4.409 (0.545)  & If morning peak & 16.658$^{**}$ (0.005) \\
If faculty & -16.881$^{*}$ (0.095) &  Distance$\times$If transit     &   3.404$^{**}$ (0.000)             \\ \bottomrule
\multicolumn{4}{l}{
\begin{tabular}[c]{@{}l@{}} Number of samples: 943; $R^2$: 0.313;  \\
$^{**}$: $p$-value $<0.05$; $^{*}$: $p$-value $<0.1$; \\
All coefficients are scaled by $1000$.\\
\end{tabular}} 
\end{tabular}

\end{table}

\begin{figure}[htb]
\centering
\includegraphics[width = 0.5\linewidth]{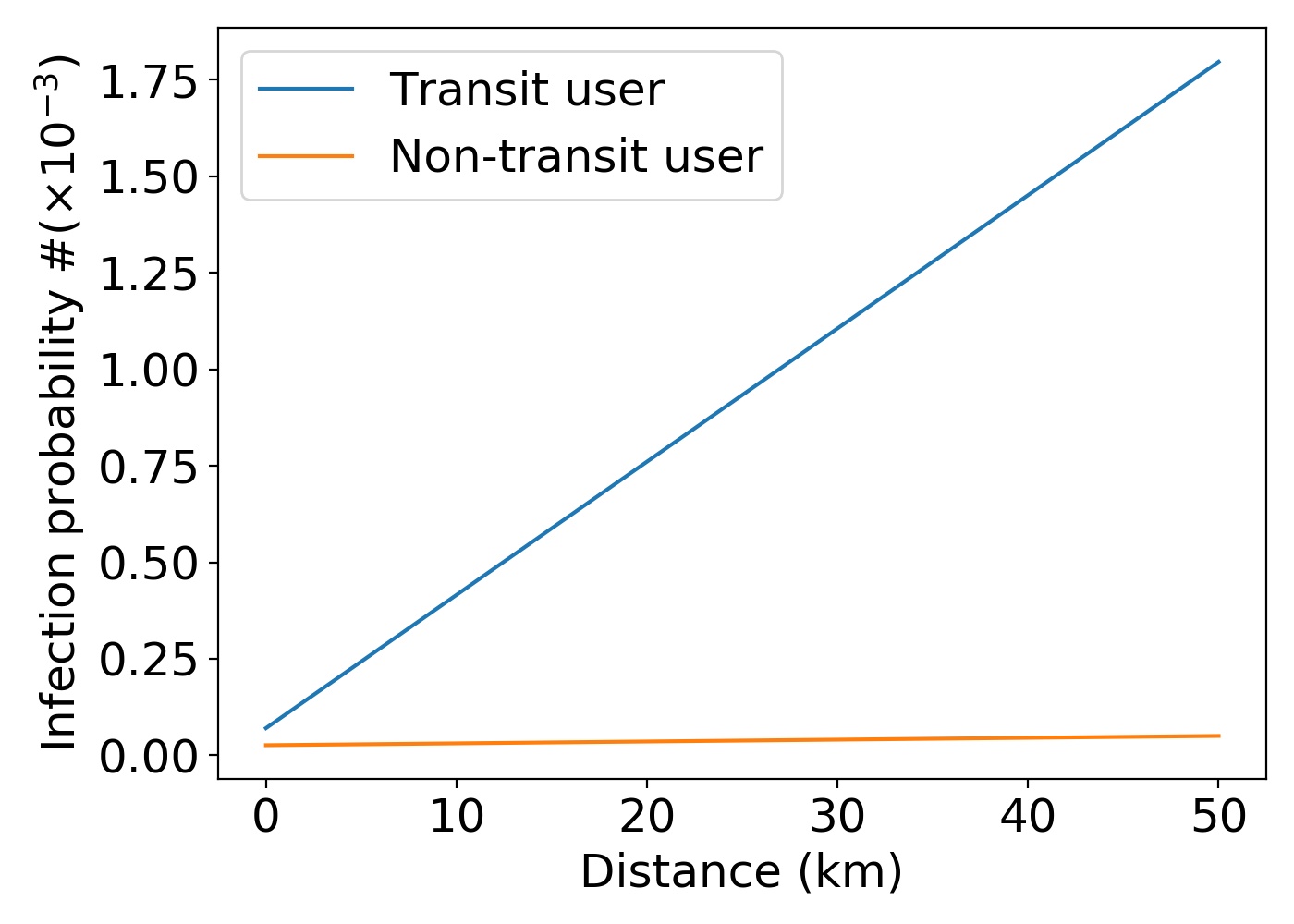}
\caption{Predicted infection probability as a function of distance}
\label{fig_distance_transit_non_transit}
\end{figure}

To better visualize the impact of distance, Figure \ref{fig_distance_transit_non_transit} shows the predicted infection probability as a function of distance for both transit and non-transit users. In general, transit users are more likely to get infected, and the infection risk increases dramatically with the increase in commute distance. However, the infection risk of non-transit users is not significantly affected by distance.

\subsubsection{Spatiotemporal distribution of infection risk}
In addition to the infection risk calculation for actual survey respondents, the proposed method can also be used to evaluate the spatiotemporal distribution of infection risks by generating synthetic observations. 

For the spatial distribution, we generate synthetic observations with home addresses in different neighborhoods. We consider two travel modes: transit and walking. Note that for dummy samples with walking travel modes, we also consider home addresses within 10km of MIT. All dummy samples' departure times are set at 8:00 AM. 

\begin{figure}[htb]
\centering
\subfloat[Transit trips]{\includegraphics[width=0.48\linewidth]{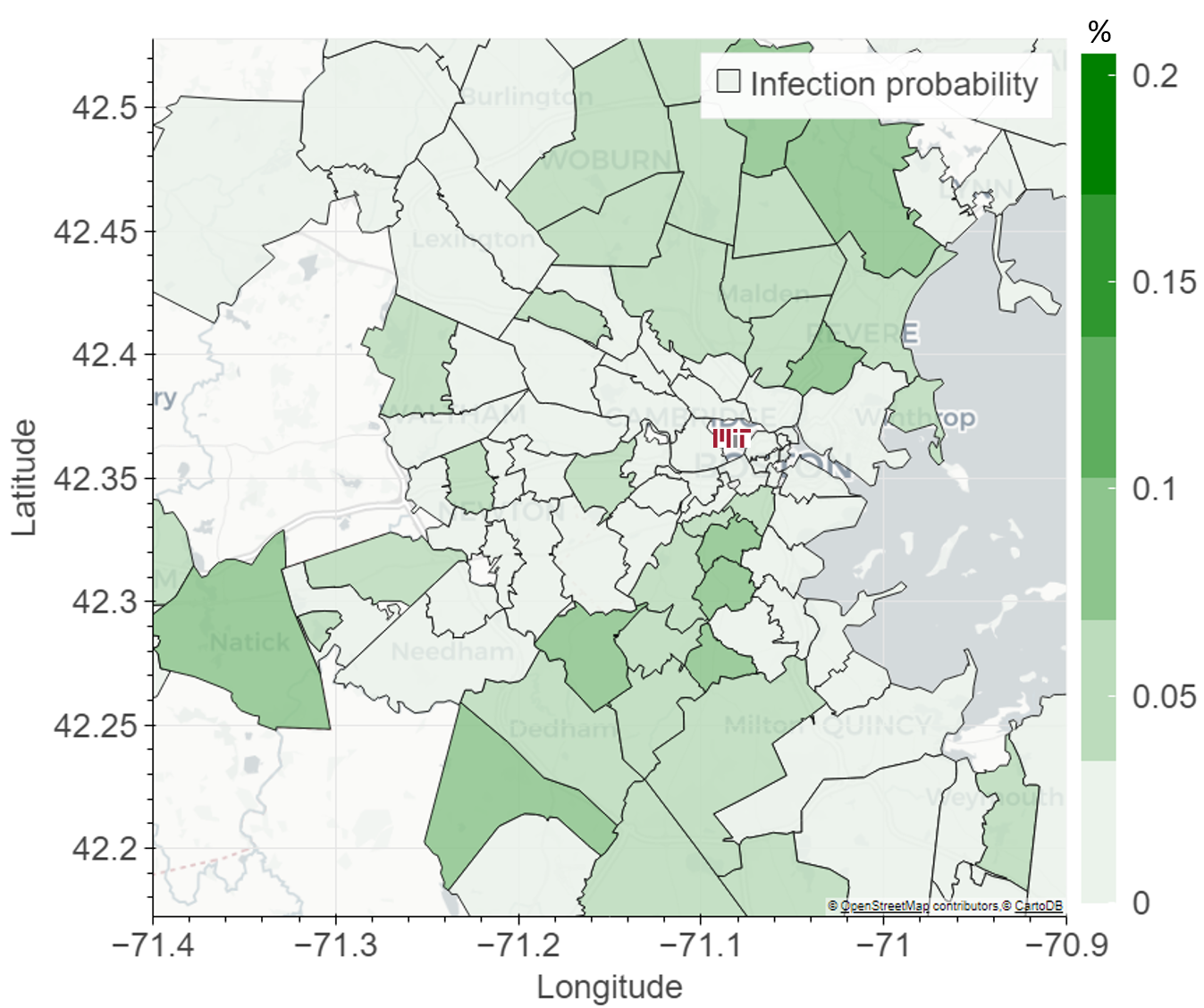}\label{fig_transit_infect}}
\hfil
\subfloat[Walk trips]{\includegraphics[width=0.48\linewidth]{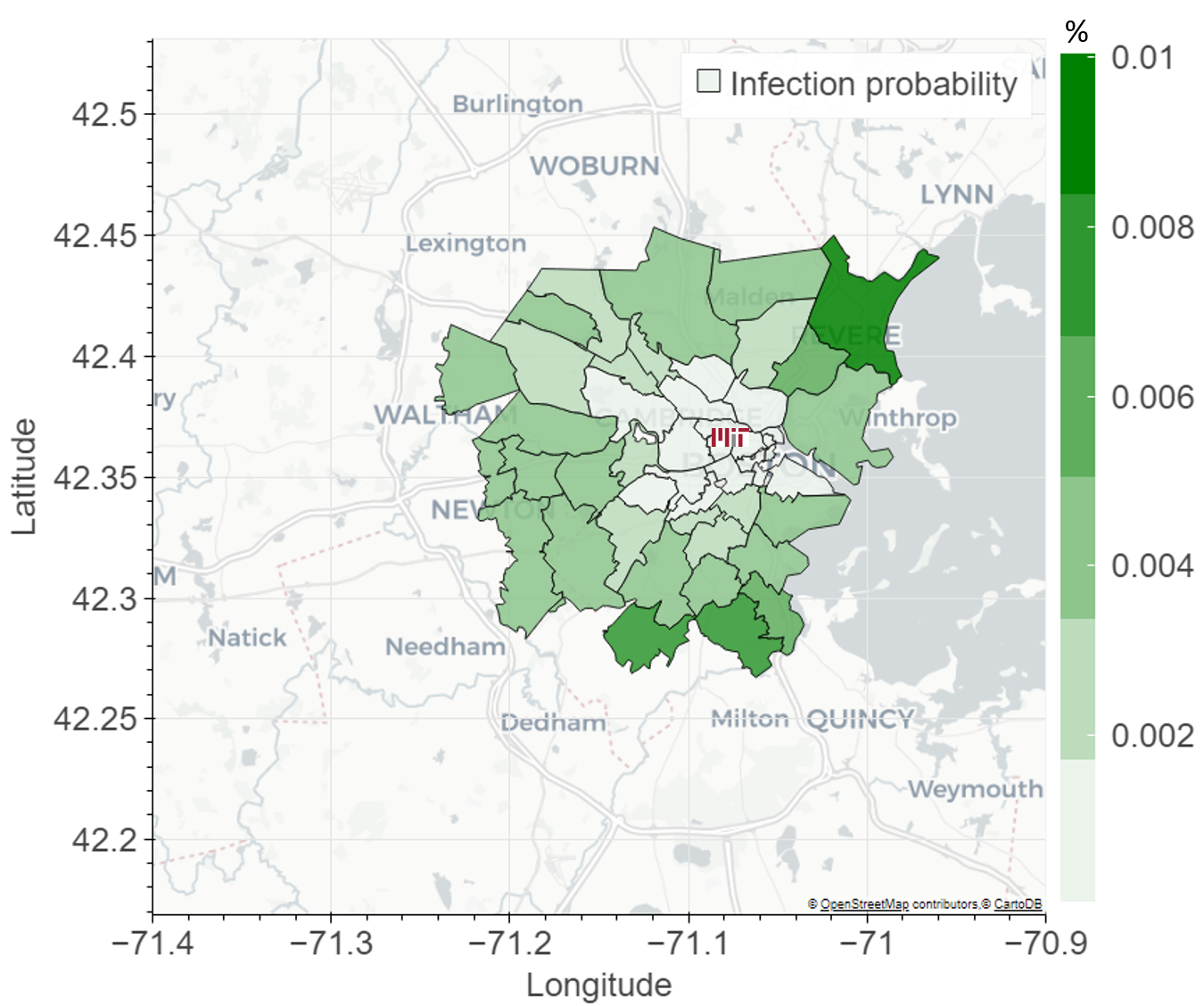}\label{fig_walk_infect}}
\caption{Spatial distribution of infection probability}
\label{fig_spatial}
\end{figure}

Figure \ref{fig_transit_infect} shows the infection probability of transit trips with home addresses in different neighborhoods. In general, people with residences further from MIT have higher infection risks. But the risk is also affected by specific transit routes (i.e., not perfectly proportional to distances). The spatial distribution for walking (Figure \ref{fig_walk_infect}) is similar. But the infection risk is much smaller than that of transit trips. 

\begin{figure}[htb]
\centering
\subfloat[Transit trips]{\includegraphics[width=0.48\linewidth]{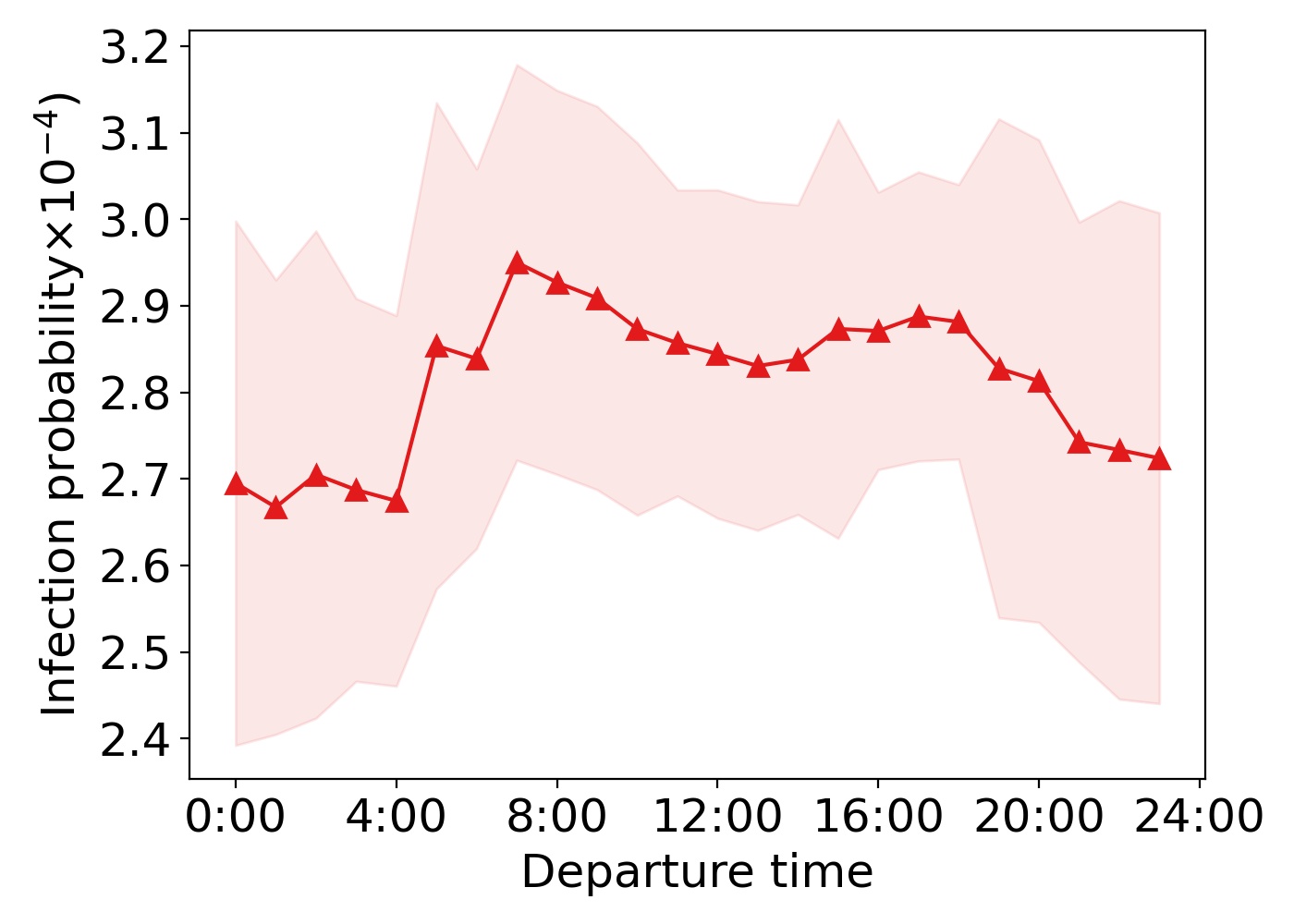}\label{fig_transit_infect_time}}
\hfil
\subfloat[Walk trips]{\includegraphics[width=0.48\linewidth]{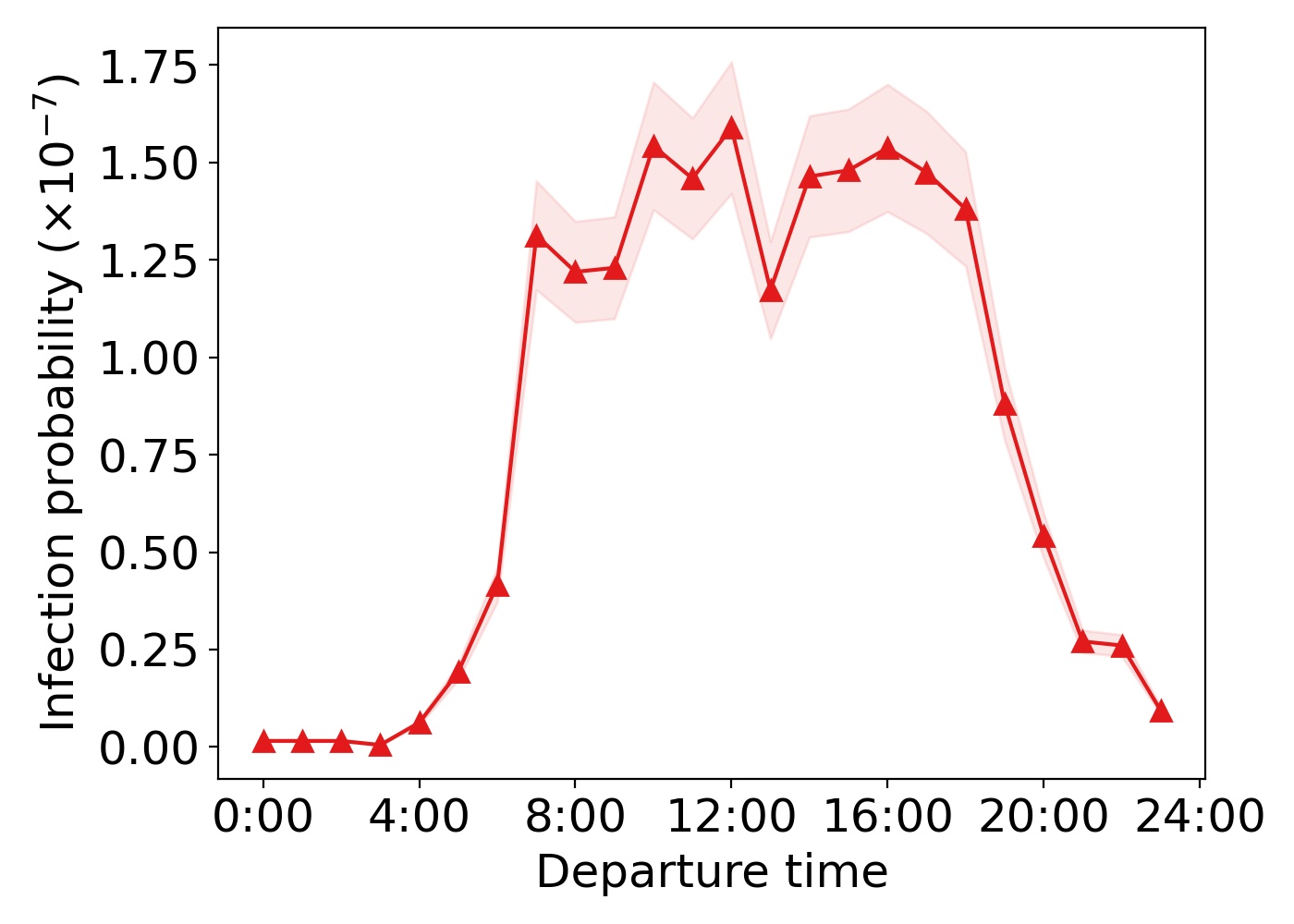}\label{fig_walk_infect_time}}
\caption{Temporal distribution of infection probability}
\label{fig_temporal}
\end{figure}

In terms of the temporal distribution, we generate synthetic observations with departure times from 0:00 to 24:00. Their home locations are assumed to be uniformly distributed across all neighborhoods. Figure \ref{fig_transit_infect_time} shows the infection risk as a function of departure time for transit trips. The shaded areas are $\frac{1}{10}$ of the estimation errors. The infection probabilities are higher when people depart during the morning and evening peak hours, which is reasonable because there is usually a higher passenger load (i.e., more close contacts) during rush hours. The infection probabilities for walking trips show a similar pattern (Figure \ref{fig_walk_infect_time}). The highest infection risks happen in the daytime (from 8:00 to 18:00), most likely due to the larger number of pedestrians. 

\section{Conclusion}\label{sec_conclusion}
The paper proposes a probabilistic framework to estimate the risk of infection during commuting considering different travel modes, including public transit, ride-share, bike, and walking. The model enables evaluating both the probability of infection and the estimation errors (i.e., uncertainty quantification). Different sources of data (such as smart card data, travel surveys, and population data) are used to estimate commuting individual's interaction with infectious environments.  The model is applied using data related to the MIT community as a case study. We evaluate the commute infection risks for employees and students. Results show that most of the individuals have very low infection probability. The maximum infection probability is around 0.8\%. Individuals with larger travel distances, traveling with transit, and traveling at peak hours are more likely to get infected. 

\begin{figure}[htb]
\centering
\includegraphics[width = 0.3\linewidth]{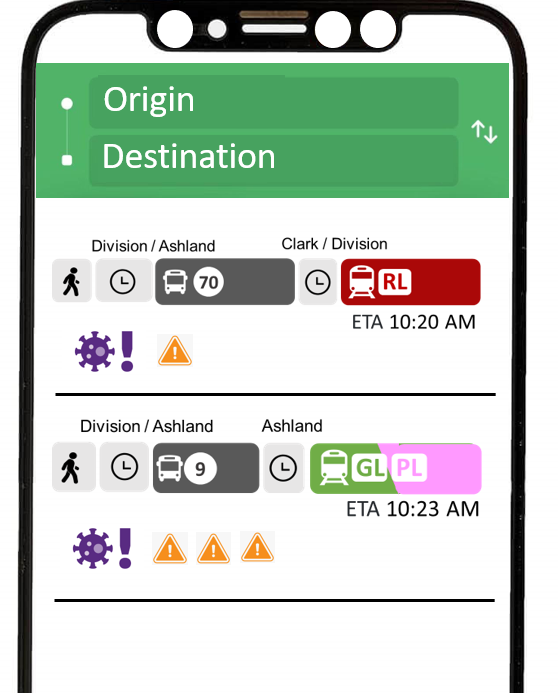}
\caption{Example of implementing the model to trip planning}
\label{fig_imp_example}
\end{figure}

The model has several practical applications. 1)  The model can be used to support decision-making for companies, schools, or communities during or post the pandemic regarding return to offices. They can collect their employees' commuting information and use the model to evaluate the commuting risk for better planning. For example, employees with high infection risk may have separate seats from the low-risk employees. 2) Another implementation of the model is to add individual-level infection risk to their trip planning tool (such as Google Map). For example, in Figure \ref{fig_imp_example}, the trip planning tool not only shows the recommended routes based on travel time, but also the infection risks. Individuals can make better path choice decisions with this additional information.  

\section{Acknowledgement}
The authors would like to thank the MIT Quest for Intelligence for their support and data availability for this research. 

\section{Author contribution statement}
\textbf{Baichuan Mo}: Conceptualization, Methodology, Software, Formal analysis, Data Curation, Writing - Original Draft, Writing - Review \& Editing, Visualization. \textbf{Peyman Noursalehi}: Conceptualization, Software, Data Curation.  \textbf{Haris N. Koutsopoulos}: Conceptualization, Writing - Review \& Editing, Supervision. \textbf{Jinhua Zhao:} Conceptualization, Supervision, Project administration, Funding acquisition.

\appendix
\appendixpage
\section{Pedestrian and cyclist density calculation}\label{app_pedestrain_data}
For the infection risk calculation of walking and bike trips, an important input is the number of close-contact encounters during the trip (i.e., the distribution of $|\mathcal{P}_s|$). The distribution can be obtained from the mean and variance of the number of cyclists (denoted as $C_{b,\tau}$) and pedestrians (denoted as $W_{b,\tau}$) at specific street $b$ for each time interval $\tau$. Since there are no GPS or trajectory data available for this study, we generate the distribution of $C_{b,\tau}$ and $W_{b,\tau}$ using the following method. 

First, we collect population data \citep{mass2020population} for each neighborhood (zip-code level) in the Boston metropolitan area. We use the Massachusetts Travel Survey (MTS) \citep{boston2011mpo} to calculate the trip generation rates given the population. As MTS does not include bike and walk mode share, the national household travel survey (NHTS) data \citep{nhts2017} is used to get the proportion of bike and walk trips as well as their temporal distributions using samples in Massachusetts. Combined with the trip generation rate, we simulate the number of bikes and walk trips for each neighborhood at different time intervals. Given limited actual bike and walk trips in NHTS data in Boston, it is hard to obtain the origin-destination (OD) distribution. But the distribution of travel distances and departure times can be obtained. For each bike walk trip, we generate the ``hypothetical'' trajectory as follows: 1) We first randomly sample an origin within the neighborhood. 2) Then, we sample the travel distance and departure time from the pre-defined distribution, as well as a specific direction (uniformly $0\sim 2\pi$). The travel distance and direction yield the destination. 3) We generate the trajectory (i.e., a sequence of streets) of the trip. From these trajectories, we obtain the number of pedestrians and cyclists in the street for each street $b$ and time interval $\tau$ (i.e., a sample of $C_{b,\tau}$ and $W_{b,\tau}$). This process is repeated multiple times to get $\mathbb{E}[C_{b,\tau}], \mathbb{E}[W_{b,\tau}]$ and $\text{Var}[C_{b,\tau}], \text{Var}[W_{b,\tau}]$. 

The generated walking and bike trip distributions are shown in Figure \ref{fig_bike_walk} (the figure is aggregated at the zip code level).

\bibliography{mybibfile}

\end{document}